\documentclass[%
 reprint,
 amsmath,amssymb,
 aps,
]{revtex4-2}

\usepackage{graphicx}
\usepackage{dcolumn}
\usepackage{bm}
\usepackage{hyperref}
\usepackage{lipsum} 
\usepackage{xcolor}
\usepackage{braket}
\usepackage{booktabs}

\makeatletter
\renewcommand{\doi}[1]{}     
\renewcommand{\eprint}[1]{}  
\newcommand{\issn}[1]{}    
\AtBeginDocument{%
  \newcommand{\xurl}[1]{}%
}
\makeatother

\makeatletter
\renewcommand{\selectlanguage}[1]{}
\makeatother 

\DeclareMathOperator{\arctantwo}{arctan2}

\usepackage{soul} 

\begin{document}

\preprint{APS/123-QED}

\title{Ultrafast configuration changes and anomalous diffusion of an aromatic adsorbate on rare-gas nanoparticles}

\author{Arne Morlok}
\thanks{These two authors contributed equally}
    \affiliation{Institute of Physics, University of Freiburg}
\author{Philipp Elsässer}
\thanks{These two authors contributed equally}
    \affiliation{Institute of Physics, University of Freiburg}
\author{Yilin Li}
    \affiliation{Institute of Physics, University of Freiburg}
\author{Ulrich Bangert}
    \affiliation{Institute of Physics, University of Freiburg}
\author{Felix Riedel}
    \affiliation{Institute of Physics, University of Freiburg}
\author{Tanja Schilling}
    \affiliation{Institute of Physics, University of Freiburg}
\author{Frank Stienkemeier}
    \affiliation{Institute of Physics, University of Freiburg}
\author{Lukas Bruder}
\email{lukas.bruder@physik.uni-freiburg.de}
    \affiliation{Institute of Physics, University of Freiburg}
\date{\today}

\begin{abstract}
Nanoparticles (NPs) exhibit tunable catalytic properties and serve as nanoreactors for controlled multimolecular chemistry. 
The kinetics and reactivity of such systems are critically governed by the surface binding configurations of adsorbates, their stochastic fluctuations, and the adsorbate mobility across the nanosurface. 
However, resolving these properties with sufficient structural, spatial, and temporal resolution remains a major experimental challenge. 
Here, we study phthalocyanine adsorbates on rare-gas clusters as a test case. 
By combining high-resolution two-dimensional electronic spectroscopy and molecular dynamics simulations, we reveal the configurational dynamics of the adsorbates and establish a direct relation between these dynamics and the nanoscale properties of the clusters. 
Our findings indicate sub-diffusive surface motion and trapping of the adsorbate within single surface facets. 
Such dynamical behavior seems unexpected considering the weak adsorbate–surface interaction and cluster temperatures close to the sublimation point. 
These results provide direct insight into the ultrafast binding dynamics of molecular adsorbates on nanoscale objects, 
which is critical for our understanding of the chemistry of such systems. 
\end{abstract}
\maketitle

\section{\label{sec:Intro}Introduction}
The chemistry on the surface of NPs plays an important role in emerging technologies and fundamental research\,\cite{zhang_surface-plasmon-driven_2018, potapov_physics_2021, mestdagh_cluster_1997, briant_reaction_2022, farnik_pickup_2021}. 
NPs exhibit tailored catalytic activity\,\cite{schwarz_menage-a-trois_2017}, act as efficient and robust photosensitizers\,\cite{zhang_surface-plasmon-driven_2018} and are key constituents in atmospheric and astrochemical processes\,\cite{potapov_physics_2021}. 
Their distinctive features arise from nanoconfinement effects, their large surface-to-volume ratio, and their tunable surface morphology. 
To develop a predictive understanding of the chemical activity of NPs therefore requires establishing direct links between the specific NP properties and the dynamics of adsorbed molecules. However, to resolve such functional relationships on the nanoscale remains highly challenging.

Here, we address this problem using rare-gas NPs as a model system. 
Rare-gas clusters constitute unique nanoreactors for studying isolated chemical processes under minimal disturbance from external influences\,\cite{mestdagh_cluster_1997, gutberlet_aggregation-induced_2009, briant_cluster_2000, farnik_pickup_2021, zhou_ultrafast_2023, jiang_ultrafast_2024}. 
These NPs can be regarded as a benchmark platform for investigating ultrafast configurational dynamics that govern surface chemistry at the nanoscale.
In these systems, the combination of small adsorption energies of reactants and near-sublimation temperatures of the clusters\,\cite{hansen_we_2004} suggest highly complex surface dynamics. 
Further, upon transition from clusters of heavier elements (Ar) to lighter elements (Ne), the clusters become increasingly liquid and show enhanced quantum properties\,\cite{de_boer_chapter_1957}. 
Eventually, rare-gas clusters appear typically in broad size distributions and thus varying surface morphology. 
This overall complexity is also reflected in previous studies reporting cases with high vs. low adsorbate surface mobility\,\cite{briant_cluster_2000, dvorak_spectroscopy_2012, dvorak_spectroscopy_2012-1} and vastly different surface binding behavior\,\cite{bangert_high-resolution_2022, awali_time-resolved_2021}. 

Experimentally, accessing the adsorbate dynamics on NPs remains challenging. 
On the nanoscale, most dynamics proceed on ultrafast time scales, requiring experimental techniques with femtosecond time resolution. 
Moreover, to understand the chemical reactivity of a system, thermally activated configuration changes are of central interest. 
However, ultrafast experimental techniques generally rely on optical excitation of the system\,\cite{maiuri_ultrafast_2020}, making it difficult to disentangle thermally induced and optically driven dynamics.
In addition, ensemble averaging over e.g. cluster-size distributions and different surface morphologies impedes establishing a functional relationship between the nanoscale cluster properties and the dynamics of molecular adsorbates. 
Conventional single-molecule methods, including fluorescence and scanning probe techniques, can in principle overcome ensemble averaging\,\cite{barkai_theory_2004}, but implementations with sufficient temporal resolution remain difficult\,\cite{brinks_ultrafast_2014, cocker_tracking_2016}. 
Likewise, although time-resolved diffraction imaging has advanced substantially, its spatial resolution is still insufficient to directly resolve configurational changes of adsorbates on individual NPs\,\cite{colombo_imaging_2023}.   

Two-dimensional electronic spectroscopy (2DES) provides an alternative approach to access the surface binding configurations of adsorbates and associated ultrafast dynamics. 
This method measures the frequency fluctuation correlation function (FFCF) of solutes in an environment\,\cite{do_measuring_2019} thus establishing a direct link between spectroscopic information and configurational dynamics despite ensemble dispersion\,\cite{mukamel_multidimensional_2000}. 

Here, we employ 2DES with particularly high spectro-temporal resolution to investigate the binding dynamics of phthalocyanine molecules adsorbed on Ar and Ne clusters as prototypical systems. 
By combining experiment with molecular dynamics (MD) simulations, we identify the characteristics of the adsorbate dynamics and relate them to the nanoscale morphology of the clusters.
The comparison between the heavier and lighter rare-gas elements further sheds light on how the increasing quantum character of the clusters affect their morphological properties, a relationship that is still highly debated\,\cite{berry_clusters_2013}. 

\section{\label{sec:Exp}Results and discussion}
\subsection{Extraction of configurational information}
\begin{figure}
    \includegraphics[width = \linewidth]{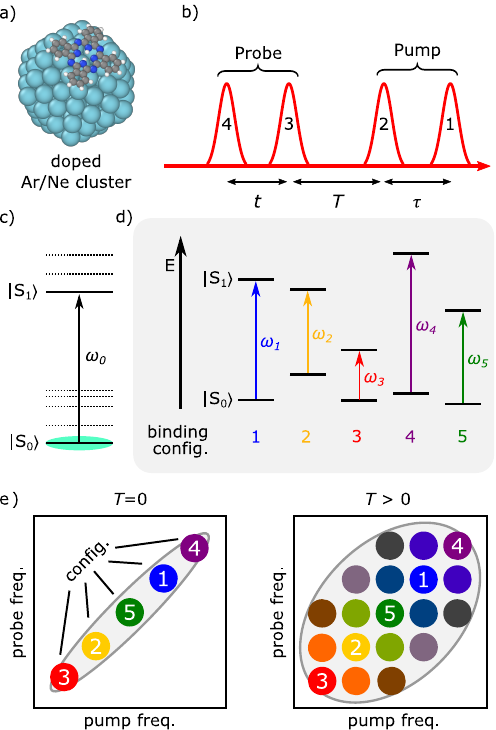}
    \caption{\label{fig:exp_scheme}Experimental scheme. \textbf{a)} Schematic of an icosahedral Ar/Ne cluster doped with a H$_2$Pc molecule. \textbf{b)} Femtosecond pulse sequence interacting with the sample. 
    \textbf{c)} Level scheme of $Q_x$-transition in H$_2$Pc. The molecule is cooled to the lowest vibrational state from where only the $0-0$ vibronic transition is accessed, both for excitation and de-excitation, due to the narrow Franck-Condon window of the vertical transition. Solid lines: vibrational ground states, dotted lines: higher-lying vibrational levels which do not contribute in the experiment. \textbf{d)} Schematic of how the ensemble of local binding configurations shift the $S_0 \rightarrow S_1$ transition inhomogeneously. \textbf{e)} 2D spectrum separating at $T=0$\,fs the inhomogeneous distribution of transition frequencies $\omega_i$ along the diagonal and the homogeneous lineprofile along the anti-diagonal direction. For $T>0$\,fs, statistical hopping between binding sites leads to a spectral broadening along the anti-diagonal direction (termed \textit{spectral diffusion})\,\cite{roberts_characterization_2006}, as a direct signature of the structural dynamics in the nanosystem.}
\end{figure}
To avoid substrate or solvent effects, we focused our study on isolated NPs in the gas phase. 
Neutral Ar and Ne clusters of diameter $< 10\,$\,nm were generated in an ultrahigh vacuum environment (see Tab.\,\ref{tab} for cluster properties and experimental details in "Methods" section). 
Evaporative cooling of the NPs results in equilibrium temperatures just below the sublimation point\,\cite{hansen_we_2004}. 
The clusters were doped with free-base phthalocyanine (H$_2$Pc) molecules (Fig.\,\ref{fig:exp_scheme}a) and interrogated upon optical interaction with the molecular  $Q_x$ transition. 
Given the low cluster temperatures and the narrow Franck-Condon window of the molecular transition, the relevant energy levels are reduced to a two-level system with negligible vibrational coupling to the cluster environment (Fig.\,\ref{fig:exp_scheme}c)\,\cite{bangert_high-resolution_2022}. 
Moreover, the $S_1$ state decays solely by fluorescence emission. 
Hence, optical excitation of the molecule is not expected to induce significant interactions with the environment. 

To extract information about binding configurations despite the ensemble averaging, we apply 2DES\,\cite{jonas_two-dimensional_2003}. 
In the condensed-phase the method is regularly implemented in a coherent four-wave mixing geometry\,\cite{fuller_experimental_2015}. 
In order to address the challenging signal-to-noise (SNR) conditions due to the low molecular density ($\approx 10^8$\,cm$^{-3}$) in the cluster-beam experiments, we implemented a highly sensitive action-detected variant of 2DES\,\cite{Bruder2019} (see also ”Methods” section). 
In this scheme, the sample is excited with a sequence of four femtosecond laser pulses and the induced fluorescence is measured (Fig.\,\ref{fig:exp_scheme}b). 
Pulse pairs $(1,2)$, and $(3,4)$ probe the time evolution of the free polarization decay between the molecular excited $S_1$ and ground state $S_0$. 
Conversely, the delay $T$ tracks the time evolution of the system in analogy to a femtosecond pump-probe experiment \,\cite{fuller_experimental_2015}. 

Fourier transform w.r.t. the coherence times $\tau, t$ yields 2D frequency maps directly correlating the excitation (pump) and detection (probe) frequencies (Fig.\,\ref{fig:exp_scheme}e). 
This correlation analysis provides high sensitivity for intra- and inter-molecular couplings and associated relaxation pathways\,\cite{brixner_two-dimensional_2005, lewis_probing_2012}. 
More importantly for the current study, photon echo-type signal contributions can be isolated such that at $T=0$\,fs, the ensemble-induced dephasing is compensated. 
Accordingly, the lineshape separates into the inhomogeneous and homogeneous broadening along the diagonal and anti-diagonal of the 2D spectra, respectively (Fig.\,\ref{fig:exp_scheme}e)\,\cite{jonas_two-dimensional_2003}. 
We use this feature to gain insight into the static properties of the nanosystems beyond the ensemble average. 

Tracking the evolution of the 2D line shapes as a function of $T$ reveals the system's FFCF\,\cite{roberts_characterization_2006, do_measuring_2019} 
\begin{equation} \label{Eq:FFCF}
    C(T) = \langle \delta \omega (T) \delta \omega (0) \rangle \, .
\end{equation} 
Here $\delta \omega(T)=\omega (T) - \langle \omega \rangle$, where $\omega(T)$ denotes the $S_0 \rightarrow S_1$ transition frequency of a specific molecule within the ensemble at time $T$ and $\langle ... \rangle$ denotes the ensemble average. 
The absence of vibrational couplings in the H$_2$Pc molecules ensures that the frequency fluctuations directly reflect changes in the adsorbate-surface binding configurations, which are activated solely by thermal energy and not by the optical interaction. 
In the case studied here, the ultrafast optical scheme thus provides a passive interrogation tool for the thermally activated structural dynamics of the nanosystems. 
\begin{table*}[]
    \centering
    \begin{tabular}{ccccccc}\toprule
         Species& $\langle N \rangle$& $T_\mathrm{cluster}$ (K) & \; 0-0 shift (cm$^{-1}$)$^\gamma$ & \; Hom. lw. (cm$^{-1}$)& $\alpha_\mathrm{exp}$ & $\alpha_\mathrm{sim}$\\ \midrule
         Ne& $3560(180)$& $10(4)$$^\beta$& $-96(12)$ & $0.45(1)$& $0.39(4)$& $0.38(1)$\\
         Ar& $253(6)$& $37(5)$$^\beta$& $-299(11)$ & $3.65(12)$& $0.28(6)$& $0.27(2)$\\
         \bottomrule
    \end{tabular}
    \caption{\label{tab}Summary of the cluster parameters along with experimental and simulation results. $\langle N \rangle$: mean number of atoms per cluster; $T_\mathrm{cluster}$: equilibrium cluster temperature; 0-0 shift: spectral shift of the $S_0,\, \nu =0 \rightarrow S_1,\,\nu' = 0$ transition; hom. lw.: FWHM value of the homogeneous linewidth; $\alpha_\mathrm{exp,\,sim}$: total diffusion paramters extracted from the experiment and MD simulation, respectively. $^\beta$ Taken from Ref.~\cite{Farges1980}. $^\gamma$ Reference energy of gas phase transition taken from Ref.~\cite{murray_visible_2011}.}
\end{table*}



\subsection{Static snapshot of the NP ensemble}
The doped NPs form an ensemble of clusters with varying sizes, surface morphologies, and adsorbate–surface binding configurations, as often encountered for nanoscale objects. 
To probe the static properties of the nanosystems, we record a snapshot of the ensemble at $T=0$\,fs. 
Using 2DES, we separate the optical response into homogeneous (single-particle) and inhomogeneous (ensemble) contributions. 
The homogeneous lineshape reflects the adsorbate–cluster interaction at the single-molecule level, whereas inhomogeneous broadening captures the distribution of structural conformers across the ensemble.

\begin{figure}
    \centering
    \includegraphics[width = \linewidth]{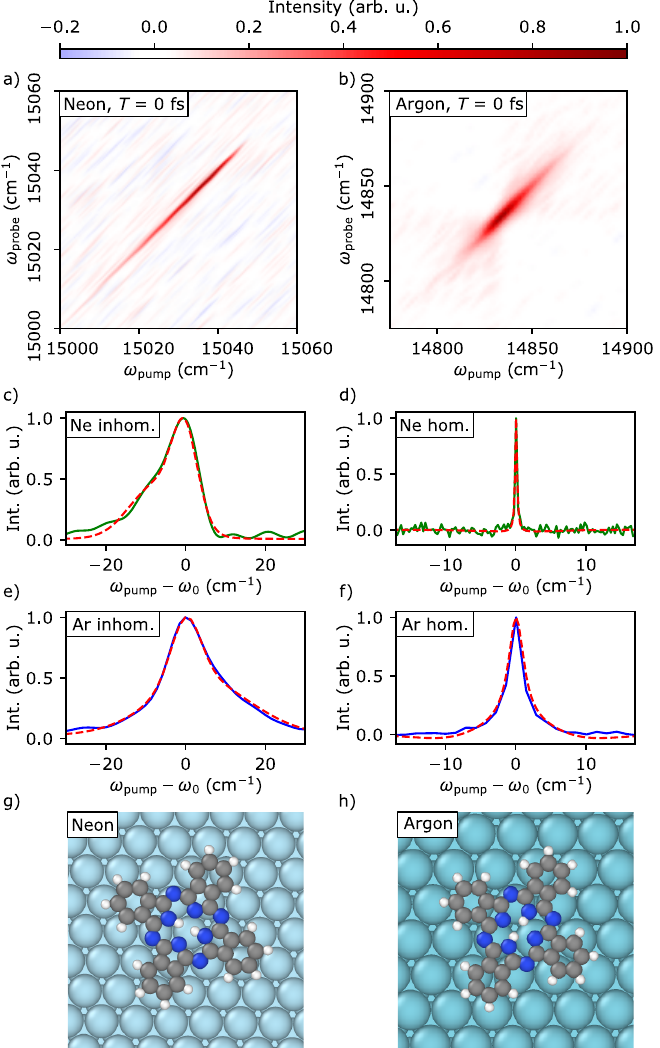}
    \caption{\label{fig:maps_and_fits} Static analysis of binding configurations. (\textbf{a, b}) 2D spectra for H$_2$Pc on Ne and Ar clusters at $T = 0$\,fs. Plotted are the real-part of the rephasing signals. (\textbf{c-f}) Homogeneous and inhomogeneous line profiles (solid) and least-square fits (dashed) extracted from averaged lineouts along the antidiagonal and diagonal of the 2D spectra and projected onto the pump axis. The lineouts are frequency shifted to overlap in the pump-axis projection. (\textbf{g}, \textbf{h}) MD simulation of H$_2$Pc on a Ne and Ar surface. The (111) surface of a face-centered cubic structure for bulk rare-gases was adopted as approximation of the facets of the icosahedral clusters.}
\end{figure}

As the light-matter interaction is reduced to the interaction with an electronic two-level system (Fig.\,\ref{fig:exp_scheme}c), the 2D spectra feature only a single peak (Fig.\,\ref{fig:maps_and_fits}a,b). 
Homogeneous and inhomogeneous linewidths are obtained from projections along the anti-diagonal and diagonal directions, respectively (Fig.\,\ref{fig:maps_and_fits}c–f). 
For quantitative analysis, we perform a global 2D lineshape fit~\cite{Bell2015} (see Supp. Info. Sec.~\ref{sec:2Dlineshape}). 
The model comprises a sum of Gaussian components along the diagonal and a Lorentzian component along the anti-diagonal. 
The excellent agreement with the data (see 1D projections in Fig.\,\ref{fig:maps_and_fits}d,f) confirms the Lorentzian lineshape of the data and thus the homogeneous character of the anti-diagonal profile. 

The extracted homogeneous linewidth (Tab.\,\ref{tab}) originates from elastic scattering of cluster phonons with the adsorbate molecule, as lifetime broadening is negligible on the relevant time scales\,\cite{bangert_high-resolution_2022}. 
Accordingly, the drastic increase in homogeneous linewidth from Ne to Ar clusters can be assigned to the greater polarizability and the higher temperature of the Ar clusters (Tab.\,\ref{tab}), increasing both the adsorbate-cluster coupling strength and the thermal population of the phonon bath. 
The stronger adsorbate-cluster coupling for Ar is also reflected in the larger environment-induced shift of the 0-0 absorption line relative to the gas-phase spectrum of isolated H$_2$Pc molecules (Tab.\,\ref{tab}). 

Within the experimental uncertainties, we find a uniform homogeneous linewidth for all diagonal positions in the 2D spectra (see residues of 2D fit in Supp. Info.\,\ref{sec:2Dlineshape}), implying that all binding sites in the ensemble couple equally strongly to the phonon bath. 
In contrast, solvation of large polyatomic molecules in bulk rare-gas matrices can lead to different solute-phonon coupling strength for distinct binding sites\,\cite{arabei_stimulated_2015, thon_wco6_2017}. 

For the inhomogeneous contribution, we observe a size dependence for small clusters and convergence of the lineprofile for larger clusters (see Supp. Info. Fig.\,\ref{fig:inhom-line_cluster_size}). 
The size-dependence is in accordance with the short-range nature of the van-der-Waals interaction, based on which convergence for cluster sizes larger than two closed shells is expected\,\cite{geissinger_hole-burning_1992}.  
In the converged regime (cluster size $\gtrsim 2500$ atoms) the inhomogeneous broadening on Ar clusters is a factor of 1.7 
larger than for Ne. 
This trend is expected given the stronger molecule-cluster coupling for Ar clusters as evidenced by the larger matrix-induced spectral shift for Ar compared to Ne (Tab.\,\ref{tab}). 
In contrast, the homogeneous lineprofile for Ar is a factor of 8.1 (Tab.\,\ref{tab}) larger than for Ne clusters and varies only by $\pm 10$\,\% with cluster size (not shown). 

Given the moderate difference in inhomogeneous broadening between the Ne and Ar clusters of same cluster size and the much broader homogeneous line broadening and greater matrix-induced shift for Ar clusters, we conclude that generally fewer adsorbate-surface binding configurations exist for the Ar clusters. 
The greater diversity of binding sites in Ne is mainly attributed to the underlying atomic lattice, featuring a finer lattice spacing thus supporting a larger number of adsorbate-binding configurations (Fig.\,\ref{fig:maps_and_fits}g,h). 

This analysis shows the benefit of capturing a static snapshot of the NP ensemble enabled by the femtosecond time-resolution of the experiment. 
The snapshot has provided us with insight into the diversity of binding sites between the two cluster species and the role of the rare-gas lattice structures.

\subsection{Stochastic molecular motion on the NPs} \label{sec:StochasticMolMotion}
\begin{figure}
    \centering
    \includegraphics[width = \linewidth]{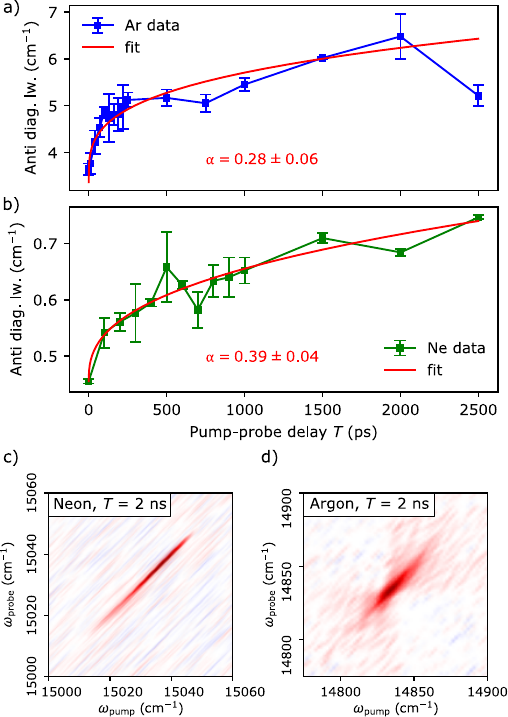}
    \caption{\label{fig:dynamics_exp_and_sim} 
    Configurational dynamics of the doped clusters. (a,b) Evolution of the antidiagonal line broadening of the 2D spectra for $H_2$Pc on Ar (blue) and on Ne (green) clusters as a function of the time delay $T$ extracted from the experimental data. The linewidth is determined from 2D lineshape fits. A fit of the data with the diffusion model (Eq.~\ref{eq:subdiff}) is shown in red. Error bars in the data reflect the standard deviation of multiple measurements for data points at $T \leq 1000$\,ps. For $T > 1000$\,ps error bars correspond the uncertainty from the 2D lineshape fit of a single dataset. The error on $\alpha$ reflect the fit uncertainty. 
    (c,d) 2D spectra of H$_2$Pc on Ne and Ar clusters at $T = 2000\,$ps.}
\end{figure}
Having established a static picture of the adsorbate–NP interaction, we now turn to its dynamical aspects. 
These are governed by thermally activated hopping between binding sites and the resulting molecular motion across the cluster surface. 
Both processes are essential to understand nanoscale chemical reactivity. 
Experimentally, it is challenging to resolve the complex configurational dynamics in an ensemble comprising a broad distribution of cluster sizes and binding sites. 
2DES provides direct access to these dynamics via the FFCF (Eq.\,\ref{Eq:FFCF})\,\cite{lazonder_easy_2006, do_measuring_2019}. 
However, very high simultaneous spectral and temporal resolution is necessary to reveal these dynamics, which is why previous studies could not observe them\,\cite{bangert_high-resolution_2022}. 
To overcome this limitation, we implemented a high-resolution variant of 2DES (see Supp. Info. Sec.\,\ref{sec:smart_scanning}) using similar sampling strategies as in frequency-comb based 2DES\,\cite{Lomsadze2018} but extended to retain temporal information.

Fig.\,\ref{fig:dynamics_exp_and_sim} shows the time-resolved data. 
Stochastic fluctuations in binding energies lead to a progressive loss of correlation between pump and probe frequencies, manifested as a time-dependent broadening along the anti-diagonal direction in the 2D spectra (Fig.\,\ref{fig:exp_scheme}e). 
The process is commonly termed spectral diffusion\,\cite{roberts_characterization_2006} and is clearly visible in the high-resolution experimental data (Fig.\,\,\ref{fig:dynamics_exp_and_sim}a,b). 
From the data we can directly conclude that thermally induced configurational dynamics occur on a $\sim 100$\,ps timescale. They are significantly faster in Ar clusters than in Ne clusters, which is consistent with the higher Ar cluster temperature. 

Furthermore, the aspect ratio between diagonal and anti-diagonal linewidth provides a measure of how much of the full configuration space is explored by each initial binding configuration. 
Full exploration would lead to a spherically symmetric 2D peak shape over time, corresponding to a complete loss of correlation between pump and probe interaction.  
In contrast, our data shows a modest anti-diagonal line broadening (compare Fig.\,\ref{fig:maps_and_fits}a,b and  Fig.\,\ref{fig:dynamics_exp_and_sim}c,d), indicating strong confinement to a configurational subspace. 

For quantitative analysis of the thermal hopping between binding sites, we apply a diffusion model\,\cite{Yuste2008,Avraham2000}: 
\begin{equation}
    \label{eq:subdiff}
   f(t)  = D\cdot t^\alpha + c,
\end{equation}
with the diffusion parameter $0 < \alpha \le 1$ and the diffusion coefficient $D$. 
Further, we add a constant offset $c$ to account for the initial homogeneous line broadening. 
For our study $\alpha$ is of main interest and the experimentally extracted values are given in Tab.\ref{tab}. 
We observe a sublinear temporal evolution ($\alpha < 1$) indicating anomalous diffusion, in contrast to normal Brownian motion ($\alpha = 1$). 
Sub-diffusive behavior is characteristic of systems with structural disorder or local trapping potentials, as for instance particles diffusing in heterogeneous media\,\cite{Yuste2008}, exciton transport in polymer chains\,\cite{maly_wavelike_2020} or structural protein dynamics\,\cite{klimek_hierarchical_2026}. 
Diffusion on the surface of NPs has been theoretically studied\,\cite{sano_theory_1981, gaigeot_diffusion_1997, fried_structure_1992}, showing indication for local trapping at surface defects\,\cite{gaigeot_diffusion_1997}. 
However, experimental evidence has been lacking  so far. 

\begin{figure*}
    \centering
        \includegraphics[width=0.9\linewidth]{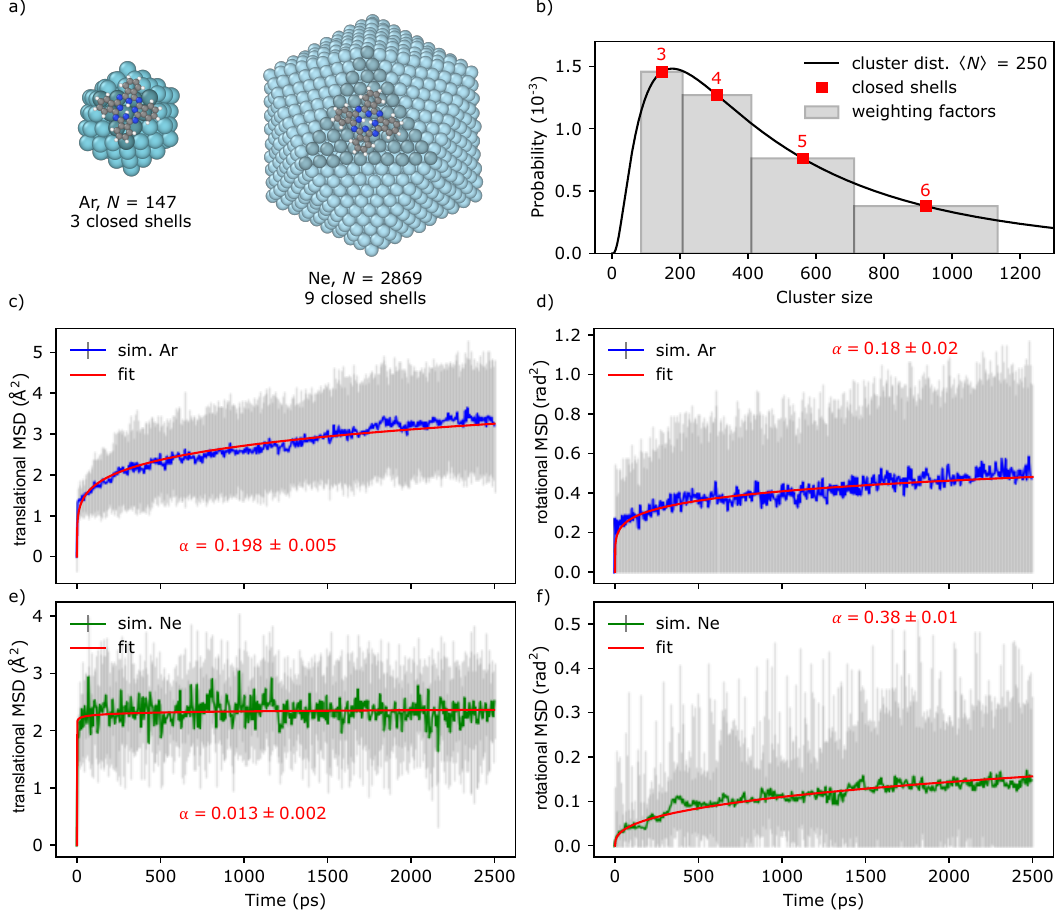}
    \caption{\label{fig:sim_results}
    MD simulations. (a) Icosahedral clusters doped with a H$_2$Pc molecule. The size of the Ar cluster ($N = 147$ atoms, 3 closed shells) and Ne cluster ($N = 2869$, 9 closed shells) is close to the most probable size in the experiment. For better visibility, one surface facet is shaded in gray. 
    (b) Ar cluster size distribution in the experiment according to $\langle N \rangle = 250$. Red squares indicate cluster sizes corresponding to 3-6 closed shells (see also Supp. Info. Fig.\,\ref{fig:SI_Ar_images}). Gray areas correspond to the weighting factors used for the averaging of cluster sizes. (c,d) MSD of H$_2$Pc on Ar clusters from MD simulation for weighted average over clusters with 3-6 closed shells. Translational and rotational coordinate in (c,d), respectively, along with least-square fits (red) of the diffusion model (Eq.~\ref{eq:subdiff}) and extracted diffusion parameter $\alpha$. Gray shaded areas indicate the simulation uncertainty. (e,f) The same for Ne clusters, however, calculated for a single cluster size of $N = 2869$. See also representative MD trajectories in Supp. Mat. video 1, 2 (Supp. Mat.\,\ref{sec:sup_mat}).}
\end{figure*}
To gain insight into the microscopic properties of the diffusion process, we performed MD simulations of the thermally activated hopping between binding sites on the cluster surface (see "Methods" section). 
The model is constructed from well-known empirical parameters (cluster temperature, particle number, first-principle/experiment based interaction potentials \cite{heinz2013thermodynamically,heinzWeb2021}) and, hence, provides an independent nanoscopic description of the studied systems. 
The only parameter adjusted manually is the interatomic binding energy $V$ between cluster atoms. 
If we use the known value from bulk rare-gas matrices, then adsorbate molecules are completely immersed in the clusters within the experimental timescale (Supp. Mat. video 1). 
This is in contradiction to experimental evidence that polycyclic aromatic molecules remain surface-solvated as can be deduced from the different absorption spectra in bulk matrices versus rare-gas clusters (compare Fig.\,\ref{fig:maps_and_fits}c,e with Ref.\,\cite{murray_visible_2011} and related experiments\,\cite{dvorak_spectroscopy_2012-1}). 
To reconcile this, we increase $V$ by a factor of 1.5, which we determined as the threshold at which full immersion is suppressed in 90\,\% of the MD trajectories (see example trajectory in Supp. Mat. video 2). 
Keeping $V$ at this threshold conserves the dynamic character of the cluster lattice which is crucial to adequately capture the thermal dynamics of a system under near-sublimation conditions. 

To map the configurational dynamics onto the experimentally observed spectral diffusion, we compute the mean-squared displacement (MSD) of the molecular hopping between binding configurations for different initial conditions. 
For the case of Ar we additionally average over the experimental cluster size distribution (mean size $\langle N \rangle = 253$ atoms), whereas for the large Ne clusters we compute a single cluster size ($N=2869$ atoms, 9 closed shells) to keep the computational effort at a feasible level (Fig.\,\ref{fig:sim_results}a,b). 

As the hopping between binding sites occurs along the translational and rotational coordinate, we computed both processes (Fig.\,\ref{fig:sim_results}c-f). 
The time evolution of the MSD in both coordinates is well described by the diffusion model of Eq.\,\ref{eq:subdiff} (see Fig.\,\ref{fig:sim_results}c-f). 
The comparison with the experiment requires convolution of the dynamics along both coordinates\,\cite{HammZanni2011}.  
Assuming Gaussian statistics, this yields
\begin{equation}
\label{eq:alpha_tot}
    \alpha_\mathrm{tot} = \sqrt{\alpha_\mathrm{trans}^2 + \alpha_\mathrm{rot}^2}\, ,
\end{equation}
where the subscripts \textit{trans}, \textit{rot} indicate the translational and rotational coordinates. 
The excellent agreement between experiment and theory (Tab.\,\ref{tab}) validates the model and enables a detailed interpretation of the sub-diffusive dynamics. 

On this basis we inspect the simulation results in more detail. 
For both cluster species, the rotational motion of the adsorbate is slower than the translation motion, which is in agreement with density functional theory calculations. These indicate that the energy barrier for the rotational coordinate on a flat surface is  approximately a factor of two higher than for the translational coordinate~\cite{michaelcomm}. 
The slower rotational dynamics in Ne compared to Ar clusters is qualitatively consistent with the lower Ne cluster temperature. 

However, the trend in the rotational dynamics is in contrast to the translational motion, which is more efficient on Ne clusters. 
This effect can be related to the larger size of the Ne clusters (Fig.\,\ref{fig:sim_results}a). 
The larger facets allow quasi-free motion within a confined region of the facet before diffusion is hindered at the facet boundaries as can be seen from the sudden steep MSD increase followed by a constant MSD level (Fig.\,\ref{fig:sim_results}e). 
In contrast, the smaller facets of the Ar clusters (Fig.\,\ref{fig:sim_results}a) impede the motion on the entire facet resulting in a gradual MSD increase (Fig.\,\ref{fig:sim_results}c). 

This relationship between adsorbate motion and cluster morphology is further corroborated by comparison of relevant length scales. 
The facet size of the icosahedral NPs closest to the maximum of the cluster size distribution are 98\,\AA$^2$ (three-shell Ar cluster) and 411\,\AA$^2$ (nine-shell Ne cluster), respectively. In contrast, the translational MSD is limited to a few \AA$^2$ in both systems. 
The larger translational MSD on the smaller Ar clusters is due to the molecule protruding the facet (see Supp. Info Fig.\,\ref{fig:SI_Ar_images}b) facilitating a larger movement of the molecule's center of mass. 

We also performed MD simulations on a large, flat face-centered cubic Ne surface (not shown). 
In this case, neither the translational nor the rotational MSD shows the spatial constraints observed in the cluster systems and the extracted $\alpha$ values are in stark difference to the experimental results. 

In total, these observations unequivocally confirm the local trapping of the adsorbates on individual cluster facets for both systems. 
This interpretation is also consistent with the large energy barriers (100–250\,meV), exemplarily determined by the MD simulations for crossing facet boundaries in the Ar clusters. 
The barriers far exceed the Ar cluster thermal energy of 3.2\,meV and hence effectively suppress inter-facet transport. 

\subsection{Conclusions}
Our study provides a microscopic, time-resolved picture of molecular dynamics on the surface of rare-gas NPs. 
This was made possible by a particularly high-resolution implementation of 2DES. 
While theoretical work has long addressed diffusion on nanoscale objects in the context of multimolecular reaction kinetics\,\cite{sano_theory_1981, gaigeot_diffusion_1997}, direct experimental access to the nanoscale dynamics has remained elusive\,\cite{briant_reaction_2022}. 
Here, we close this gap by resolving the thermally driven configurational dynamics of adsorbates at the single-particle level and by comparing these results with a computational study.

By isolating the relevant dynamical contributions, we uncover local spacial trapping of the adsorbates within individual cluster facets, 
leading to sub-diffusive molecular motion. 
This finding demonstrates that even in seemingly weakly bound, thermally activated environments, the local structure of the NP surface imposes strong constraints on molecular motion and thus the multimolecular chemistry. 
The results further establish the solid-like surface character of the studied Ar and Ne rare-gas cluster, and resolve the debate of whether liquid-like surface layers are present under these conditions\,\cite{berry_clusters_2013}.

More broadly, 
our work provides a mechanistic framework for adsorbate dynamics on nanoscale systems and establishes a direct link between these dynamics and the surface properties of the NPs. 
The applied experimental configurational analysis has been also employed in complex condensed-phase systems and environments with strong spectral broadening\,\cite{moca_two-dimensional_2015, sanda_center_2015, do_measuring_2019, yoshida_anomalous_2020}. 
Accordingly, studies of adsorbate dynamics on other NPs such as metal and metal-oxide NPs are within reach, with direct relevance to catalysis and solar energy conversion\,\cite{schwarz_menage-a-trois_2017, zhang_surface-plasmon-driven_2018}.

\section{Methods} \label{sec:methods}
\subsection{Sample preparation}
Details about the vacuum apparatus to prepare the rare-gas cluster beams can be found in Refs.\,\cite{Bruder2019, bangert_high-resolution_2022}. 
The general technique of doped rare-gas cluster spectroscopy is reviewd in Refs.\,\,\cite{mestdagh_cluster_1997, toennies_superfluid_2004}. 
Briefly, the clusters are generated upon continuous gas expansion into vacuum with parameters: 10\,\textmu m nozzle orifice diameter, 296\,K nozzle temperature, 90\,bar stagnation pressure for Ar cluster generation and 5\,\textmu m, 65\,K, 50\,bar for Ne cluster generation. 
We performed polarization anisotropy dephasing measurements (similar to Ref.\,\cite{Awali2014}) for an accurate estimate of the corresponding mean cluster sizes and obtained $\langle N_\mathrm{Ar} \rangle \approx 250$ for Ar and $\langle N_\mathrm{Ne} \rangle \approx 3600$ for Ne. 
We note that this approach is more accurate than the empirical Hagena scaling law for cluster size estimation\,\cite{Hagena1972,Hagena1981}, which tends to underestimate the size of larger clusters due to fragmentation in the detection scheme used to derive the scaling law (details will be published elsewhere). 

After passing a skimmer (opening diameter of 400\,\textmu m) the cluster beam is doped upon passing through a H$_2$Pc vapor cell of 10\,mm length. 
The vapor pressure is controlled by radiative heating of H$_2$Pc powder (Sigma Aldrich, 29H,31H-Phthalocyanine, 98\,\%) up to 365\,$^\circ$C. 
The doped cluster beam travels through a 5\,mm orifice into the adjacent detection chamber to suppress contributions from free H$_2$Pc molecules emitted effusively from the vapor cell. 

Cluster temperatures are given in Tab.\,\ref{tab}, which correspond to thermal energies of $0.7\,$meV (6\,cm$^{-1}$) for the Ne and $3.2\,$meV (26\,cm$^{-1}$) for the Ar clusters. 
In comparison, the lowest active vibrational mode of H$_2$Pc has an energy corresponding to $90\,$cm$^{-1}$, leading to Boltzmann factors of $2\times 10^{-6}$ for Ne and $3\times 10^{-2}$ for Ar, thus justifying the assumption of vibrationally cold molecules. 

\subsection{Spectroscopic method}
The 2DES is performed in a collinear beam geometry using action-detection. 
Details can be found in Ref.~\cite{Bruder2019}. 
The collinear four-pulse train is focused into the vacuum apparatus and the induced fluorescence is imaged perpendicular to the laser beam propagation direction onto a photo multiplier tube using a large numerical aperture lens. 
The output signal of the detector is fed into the current input of a digital lock-in amplifier. 

The laser parameters for the Ar and Ne cluster measurements are central wavenumber: 14825\,cm$^{-1}$ and 14925\,cm$^{-1}$, spectral full width at half maximum: 600\,cm$^{-1}$ and 575 \,cm$^{-1}$, respectively. 
Further parameters are laser repetition rate: 200\,kHz, typical pulse length: $\approx 50$\,fs, pulse energy: $55\,$nJ and focus diameter: $150$\,\textmu m, all defined at the interaction volume in the vacuum apparatus. 
The cluster beam travels at a velocity of $\approx 500$\,m/s. 
Given the laser focus diameter and repetition rate, this ensures renewal of the sample in-between each laser cycle. 
Hence, photo bleaching and accumulation effects over multiple laser cycles are avoided despite the long data acquisition times in the experiment. 
The background contribution of hot H$_2$Pc molecules effusively emitted from the doping cell are removed by differential measurements with and without blocked cluster beam. 

The fluorescence amplitude is detected both for rephasing and non-rephasing signal contributions (details below) as a function of the pulse delays $\tau, T$ and $t$ to construct 2D maps spanned by the $(\tau, t)$ tuple and $T$ acts as a parameter. 
Zero-padding is applied to the data along the $\tau$ and $t$ dimension.  
A Fourier transform w.r.t. $\tau, t$ yields the 2D frequency spectra and $T$ describes the time evolution of these spectra. 
Due to the large aspect ratio between homogeneous and inhomogeneous line broadening, the non-rephasing signal contribution is negligibly small. 
Therefore, we evaluate the real part of the rephasing Fourier spectra, which is shown in Figs.\,\ref{fig:maps_and_fits}, \ref{fig:dynamics_exp_and_sim} of the main text. 

The number density of the doped clusters in the interaction volume of the experiment is $\approx 10^8$\,cm$^{-3}$, posing a sensitivity challenge for the experimental method, which is solved by combining rapid phase modulation of the optical pulses with efficient lock-in detection\,\cite{tekavec_fluorescence-detected_2007, Bruder2019}. 
To this end, the carrier-envelope phase of each pulse is modulated at a frequency of $\Omega_i$ $(i = 1-4)$ using acousto-optical modulators. 
This induces a frequency beating at the difference frequencies $\Omega_{21}=5$\,kHz and $\Omega_{43}=8$\,kHz between the electric fields of pulses (1,2) and (3,4), respectively.
Coherent excitation with the modulated four-pulse train induces a characteristic modulation of the fluorescence yield, enabling facile separation of linear background and the nonlinear  signals. 
While the linear background appears at DC frequencies and modulations of $\Omega_{ij}$ ($i,j = 1-4$), the non-linear 2DES signal is detected at the mixing frequencies $\Omega_{21} - \Omega_{43}=3$\,kHz for the rephasing and $\Omega_{21} + \Omega_{43} = 13$\,kHz for the non-rephasing signal. 
The lock-in amplifier filters on the rephasing and non-rephasing signal modulation and thereby suppresses the background with very high contrast, which yields the high dynamic range required for the experiment. 
This is combined with phase-sensitive lock-in amplification using as reference the optical four-pulse interference spectrally filtered in a monochromator. 
The comparative phase measurement between 2DES signal and optical reference results in cancellation of most of the phase noise from the optical interferometers. 

The low target density requires relatively long signal integration times of $\sim 2$\,s per $(\tau, T, t)$ data point, despite the efficient signal-recovery technique. 
In addition, the very high spectral resolution of the measurements implies long $\tau, t$ scan ranges and thus acquisition of many data points. 
For the Ar, Ne cluster measurements, $\tau, t$ are scanned 0 to 16\,ps and 0 to 100\,ps in 160\,fs and 320\,fs step increments for a single static spectrum, respectively. 
These spectra are recorded for a $T$ range from 0 to 2500\,ps. 
Recording such large 3D data arrays at the given integration time requires very long data acquisition times, rendering it impossible to map the ultrafast dynamics of the target systems at full spectral resolution\,\cite{bangert_high-resolution_2022}.  
In order to solve this problem, we developed an approach which drastically reduces the required amount of data points 
without loss of resolution (details in Supp. Info. Sec.\,\ref{sec:smart_scanning}). 

\subsection{\label{sec:comuptational} Computational methods}
The dynamics of the H$_2$Pc molecule on the cluster surfaces is investigated by means of MD simulations. We describe the pair interactions between the atoms with the PCFF-IFF \cite{heinz2013thermodynamically,pramanik2017carbon,heinzWeb2021}, an all atom force field that was build to simulate inorganic–organic interfaces. The non-bonded pair interactions are described by a combination of 9-6-van der Waals terms, and for (partially) charged atoms, Coulomb contributions. The charges are assumed to be point charges at the sites of the atoms. The rigidity of the H$_2$Pc molecule is controlled with respect to bond lengths, angles between bonds, and torsion angles, as well as cross terms between these. The parameters for the interactions are those published by Heinz et al. \cite{heinzWeb2021}. Their applicability to similar systems has been demonstrated previously \cite{deshpande2021prediction,elsasser2023optimizing,elsasser2024structural,patil2025high}. Our only deviation from these interaction parameters is in the pair interaction of the rare-gas atoms. Due to their origin in bulk measurements, we increase the pair interaction strength in the clusters to $V=1.5V_0$. This is chosen such that the surface tension is just strong enough to support the H$_2$Pc molecule in order to prevent the molecule from sinking into the cluster (see main text and Supp. Material video 2, 3). 

The system is set up with a H$_2$Pc molecule in close proximity to the rare-gas surface. For the configuration of the rare-gas atoms, we utilize two different setups: 

(a) In order to determine the motion on a non-constrained substrate, we set the atoms in a face-centered cubic lattice. A surface is created by leaving a three atom thick layer at the bottom $xy$ plane of the simulation box with periodic boundary conditions in $x$ and $y$ direction. Depending on the simulation, the orientation of the surface is either $(001)$ or $(111)$. 

(b) For the major part of the simulations, we are interested in the constrained surfaces of atomic clusters. These clusters are initially set up as icosahedrons. This is based on the observation that this is the energetically preferred configuration for Lennard-Jones particles, also if the cluster is liquid just above the melting point or supercooled \cite{vsiber2004vibrations,elsasser1987energetics}.  The temperatures, which are controlled by a Nose-Hoover thermostat \cite{nose1984unified,hoover1985canonical}, are chosen to fit the experimental conditions of $\Theta_{\mathrm{Ar}}=37$\,K for Ar and 10\,K for Ne clusters.

The system is equilibrated for at least 7000\,ps, before we start investigating the spatial and angular movement of the molecule. The position of the H$_2$Pc molecule is defined by its center of mass, relative to the surface of the cluster. In order to correct the motion of the molecule for the changes in the cluster-configuration, we subtract the average motion, both rotational and lateral, of the cluster from the movement of the molecule. 

For Ar clusters, simulations are performed for cluster sizes corresponding to 3-6 fully closed shells (Fig.\,\ref{fig:SI_Ar_images}, Tab.\,\ref{tab:cluster_size}) and the result is averaged using the weights from the log-normal size distribution (see main text, Fig.\,\ref{fig:sim_results}b). 
For the much larger Ne clusters, computational effort is considerably larger and only one cluster size corresponding to nine closed shells is simulated. 
This size is closest to the most probable cluster size of the log-normal distribution for the Ne cluster experiments. 
For each cluster size, calculations of 100 independent trajectories with a time evolution of $T=2500\,\text{ps}$ in $2.5\cdot10^7$ increments is performed. 
Within the above defined simulation parameters, we do not observe a lateral diffusion over multiple facets and we can define the translational MSD as  
\begin{align}
    \braket{r^2(T)} &=\braket{|\vec{r}_i(T)-\vec{r}_i(0)|^2} \nonumber\\
    &= \dfrac{1}{N}\sum_{i=1}^N |\vec{r}_i(T)-\vec{r}_i(0)|^2 \label{eq:msd}
\end{align}
with the number of trajectories $N$, the center-of-mass position at time $T$, $\vec{r}_i(T)$, and the initial, equilibrated position $\vec{r}_i(0)$. 

The MSD for the angular motion (rotational MSD) is determined analogously: The orientation of the molecule is defined as the axis through the two NH groups in the H$_2$Pc molecule. We construct with the Gram-Schmidt process a reference frame, which is based on the orientation of the H$_2$Pc molecule at $T=0$ ($x$ axis) and a vector that stands orthogonal on the molecule ($z$ axis), as well as a vector that is orthogonal to the two previous ones ($y$ axis). In the case of non-planar motion of the molecule, e.g. by large lateral motion, this reference frame would still be valid by rotating the coordinate system with the motion of the molecule on the cluster-surface, but not with the local rotational motion of the molecule. With this transformation, we can determine the angular deviation from the initial orientation by the spherical coordinate
\begin{align}
    \varphi(T) = \arctantwo(x(T),y(T)).
\end{align}
$x(T)$ and $y(T)$ are the corresponding coordinates of the orientation in the new coordinate system at time $T$. The rotational MSD is thus defined by
\begin{align}
    \braket{\varphi^2(T)} =  \dfrac{1}{N}\sum_{i=1}^N \varphi_i^2(T). \label{eq:msad}
\end{align}
The definition range of $\braket{\varphi^2(T)}$ is restricted to $[0,\pi^2]$. This is motivated by the experimentally measurable configuration space: due to the symmetry of the H$_2$Pc molecule, no additional distinct configurations exist for larger rotations. 

The averaged translational and rotational MSD traces along with their uncertainties are shown in Fig.\,\ref{fig:sim_results}. The uncertainties are in both cases obtained by calculating the standard deviation of subsets of the trajectories. As a result of the limited number of trajectories, this results in an overestimation of the uncertainty.

As an additional test, we determine an estimate for the energy barriers between the facets of the icosahedrons (see main text Sec.~\ref{sec:StochasticMolMotion}). For this, we assume an equilibrated atomic cluster to be rigid. The molecule is coupled to a particle that does not interact with the system via a soft spring. By rotating this particle around the cluster on different trajectories, the molecule is forced to move over the edges of the atomic cluster.

\section{Funding}
Funding from Deutsche Forschungsgemeinschaft RTG 2717. This work is also based upon the work of COST Action CA21101 "Confined molecular systems: from a new generation of materials to the stars" (COSY) supported by COST (European Cooperation in Science and Technology). The authors acknowledge support by the state of Baden-Württemberg through bwHPC and the German Research Foundation (DFG) through Grant No. `INST 39/963-1 FUGG' (bwForCluster NEMO), as well as `INST 39/1232-1 FUGG' (bwForCluster NEMO 2).

\section{Acknowledgment}
We acknowledge fruitful discussions with Bernd von Issendorff. 

\section{Disclosures}
The authors declare no conflicts of interest.

\section{Data Availability Statement}
  The experimental and simulation data included in this work are available on the open repository: \textit{Accession codes will be available before publication}.

\newpage
\clearpage
\appendix

\section{Supplementary information}
\subsection{\label{sec:smart_scanning}Reduction of data points for high-resolution 2DES}
\begin{figure}
    \centering
    \includegraphics[width = \linewidth]{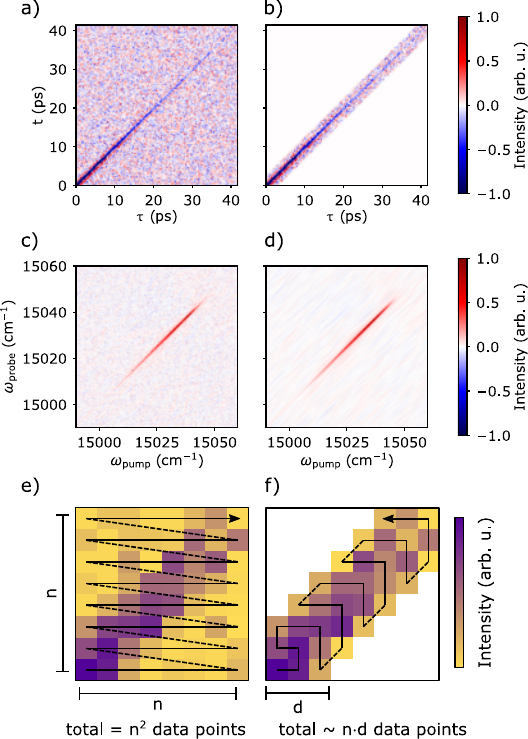}
    \caption{\label{fig:diag_scanning} Data sampling scheme to reduce acquisition time for high-resolution 2DES. \textbf{a} Time-domain 2D map (real-part of the rephasing signal) recorded for H$_2$Pc attached to Ne clusters. The signal amplitude is concentrated around the diagonal of the 2D map. 
    The majority of off-diagonal data points contain no signal and are dominated by noise. 
    \textbf{b} Same data set as in (a) but cropped along the antidiagonal dimension. In this way obsolete data points only containing noise are omitted. 
    \textbf{c,d} Fourier-transforms of the 2D maps shown in a and b, respectively. In both cases the same spectral profile is obtained, however, with reduced noise in the latter case. 
    \textbf{e} Schematic of line-by-line acquisition of a 2D map in the time domain. The $\tau$-delay stage is scanned for $n$ increments. After each scan, the stage returns to $\tau=0$ and $t$ is incremented by one step. Solid/dashed black lines indicate stage movement while acquisition is active/paused, respectively. \textbf{f} Reduced data sampling scheme. Same notation as in (e).}
\end{figure}

In action-detected 2DES, the data is recorded in the time-domain, yielding 2D data maps as a function of $(\tau, t)$ with $T$ being a parameter that is incremented for each new 2D map. 
The $(\tau,t)$-matrix is commonly acquired line by line. 
However, depending on the spectral response of the sample, the amplitude distribution in the time-domain data can be rather sparse and hence many data points are recorded which contain no information. 
In our study, the measured samples exhibit a large aspect ratio between the inhomogeneous and homogeneous linewidths. 
Accordingly, in the time domain the signal is concentrated around the diagonal of the 2D maps rendering a large portion of the off-diagonal data points obsolete (Fig.\,\ref{fig:diag_scanning}a). 

We solve this issue with a coordinate-transform $(\tau, t) \rightarrow (\tau', t') = (\tau+t, \tau - t)$ which effectively rotates the coordinate system by $45^\circ$. 
This coordinate-transform disentangles the homogeneous and inhomogeneous line broadening in the time domain: the former broadening appears along the $\tau'$, the latter along the $t'$-axis. 
The large ratio between homogeneous and inhomogeneous broadening translates to a large ratio between the coherence times along the $\tau'$- and $t'$-axis.  
In particular, due to the short coherence time along the inhomogeneous broadening dimension, data acquisition along the $t'$-axis can be apodized after a few ps without loss of spectral resolution (Fig.\,\ref{fig:diag_scanning}a-d). 
In this way, the amount of required data points can be drastically reduced and very high resolution 2D spectra can be recorded within reasonable acquisition times. 

In practice, we implement this acquisition scheme by recording the time domain data only around the diagonal. 
Instead of recording a $n\times n$ matrix for the $(\tau, t)$-delays (Fig.\,\ref{fig:diag_scanning}e), we record a reduced matrix of $d \times d$ size and increment this matrix along the diagonal direction of the time-domain 2D map (Fig.\,\ref{fig:diag_scanning}f). 
In this way, the amount of required data points reduces from $n^2$ to $n^2-(n-d+1)(n-d) = 2nd + d - d^2 - n \propto n\cdot d$. 

We note, that slightly more data points could be omitted by physically implementing the $(\tau, t) \rightarrow (\tau', t')$ coordinate-transform in the experimental setup. 
However, since this requires changes in the optical setup and the gain is not drastic in our case, we refrained from this step. 
We further note, that physical implementation of such coordinate-transform has been demonstrated in frequency-comb based 2DES\,\cite{Lomsadze2018}. 
In this scheme the delay $T$ could not be scanned due to which only static 2D spectra can be recorded, which can be sufficient in many applications. 
However, in our study information about the dynamic evolution of the system is essential to reveal the nanoscopic properties of the system. 

\begin{figure*}
    \centering
    \includegraphics[width = 0.9\linewidth]{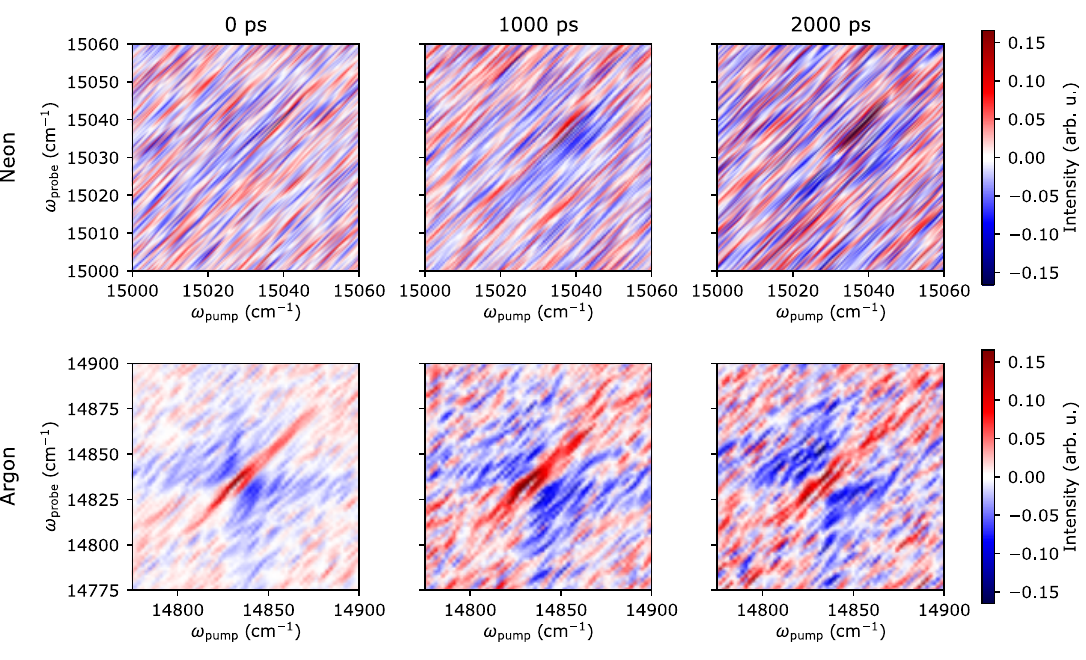}
    \caption{\label{fig:SI_residuals} 2D maps of the residuals between the 2D lineshape fit and the experimental data of H$_2$Pc on Ne and Ar clusters at waiting time delays of $T = 0$\,ps, $T = 1000$\,ps, and $T = 2000$\,ps.}
\end{figure*}
\subsection{\label{sec:2Dlineshape}Extraction of the homogeneous linewidth}

In overall, this effect may lead to a slight overestimate of the antidiagonal linewidth extracted from the 2D lineshape fit. 
Since this is a systematic error, it does not affect the diffusion dynamics analyzed in the main text. 
To extract the homogeneous linewidth from the 2D spectra, we performed a 2D lineshape fit adopting the model in Ref.~\cite{Bell2015}. 
Our fit function comprises of a single Lorentzian profile along the antidiagonal dimension and a sum of Guassian functions along the diagonal dimension. 
The latter is chosen to adapt to the structured inhomogeneous lineprofiel of the samples. 
We note that this choice does not influence the homogeneous linewidth but improves the quality of the fit. 
The resulting 2D lineshape function used for the fit is
\begin{align}
    S_R(\omega_t,\omega_\tau) =& \sum_i^{1,2} A_i \dfrac{1}{2\sigma_i(2\gamma - i(\omega_t + \omega_\tau))} \\ \nonumber 
    & \times \left\{ e^{\frac{(\gamma -i(\omega_t - \omega_{0i}))^2}{2\sigma_i^2}} \mathrm{Erfc}\left[ \dfrac{\gamma - i(\omega_t - \omega_{0i})}{\sqrt{2} \sigma_i} \right] \right.  \\ \nonumber
    & + \left. e^{\frac{(\gamma -i(\omega_\tau + \omega_{0i}))^2}{2\sigma_i^2}} \mathrm{Erfc}\left[ \dfrac{\gamma - i(\omega_\tau + \omega_{0i})}{\sqrt{2} \sigma_i}\right]  \right\}.
\end{align}
Here, $A_i$, $\omega_{0i}$ and $\sigma_i$ refer to the amplitudes, center frequencies and widths of the Gaussians. The dephasing rate $\gamma$ is connected to the full-width at half-maximum (FWHM) of the homogeneous broadening by $\textrm{FWHM} = 2\gamma$.\\ 

Lineshape analysis is commonly performed in the total 2D spectra, which is a sum of the rephasing and non-rephasing 2D spectra. 
However, due to the particular narrow homogeneous linewidth in our study, the non-rephasing signal is orders of magnitude weaker than the rephasing signal. 
Hence, analyzing the total 2D spectra is unfavorable concerning the noise introduced by the weak non-rephasing signal. 
Instead we extracted the homogeneous lineshape directly from the rephasing part of the 2D spectra. 
The good agreement between the model and the data is visible in the 1D projections of the data and fit (see main text Fig.~\ref{fig:maps_and_fits}). 
In addition, Fig.~\ref{fig:SI_residuals} shows residuals between the 2D rephasing spectra and the fit.  
For the Ne clusters, we obtain unstructured residuals, indicating absence of systematic deviations between the data and the fitted model. 
In contrast, for the Ar clusters, a cross-like structure with a elongated peak of positive amplitude in the center and wings of negative amplitude can be seen. 

This pattern arises from an aliasing artifact from the Fourier transform of the time-domain data. In the data, wings from the center peak that extend beyond the sampled frequency window are aliased back into the spectrum and compete with the expected negative features for the rephasing signal. This aliasing artifact also explains the positive residual at the position of the peak in the data. 


The 2D lineshape fits are based on a model assuming a single homogeneous linewidth for all binding configurations. 
In principle, the different adsorbate-cluster binding configurations could lead to different coupling strength with the phonon bath and thus to a varying homogeneous linewidth along the diagonal of the 2D spectra. 
However, analyzing the residuals between the 2D linshape fit and the data (Fig.~\ref{fig:SI_residuals}) does not point to multiple homogeneous linewidths within the ensemble. 
Since no systematic deviations between the fit and the data along the diagonal of the 2D spectra is observed, we can conclude, that all binding configurations exhibit the same coupling strength to the phonon bath, at least within the experimental resolution. 

\subsection{\label{sec:inhom_linewidths}Dependence of the inhomogeneous lineprofile on the mean cluster size}
\begin{figure*}[b]
    \centering
    \includegraphics[width=0.8\linewidth]{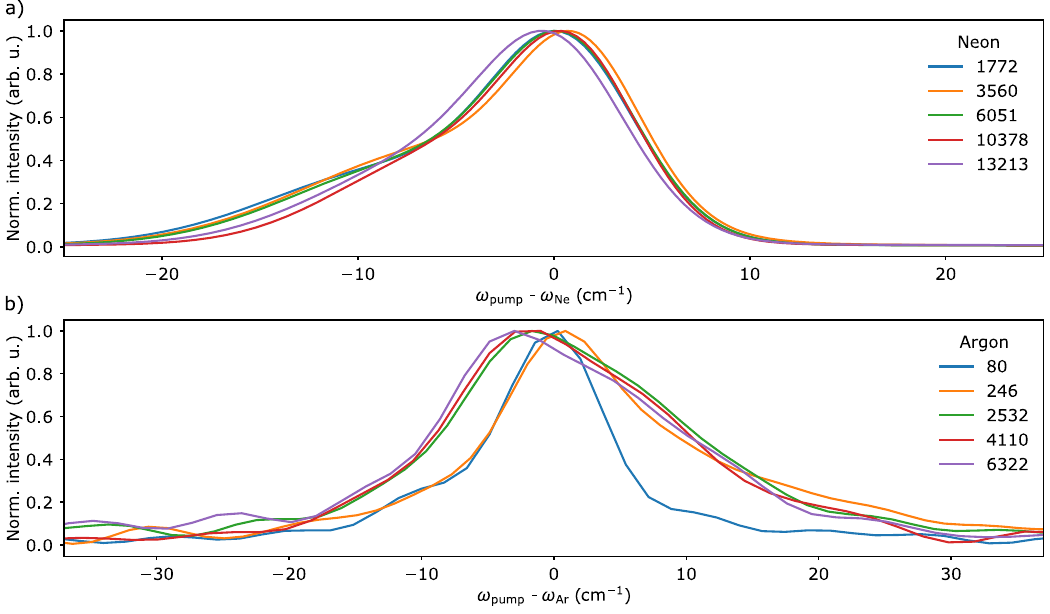}
    \caption{\label{fig:inhom-line_cluster_size} Inhomogeneous line profile of H$_2$Pc on Ne (\textbf{a}) and Ar (\textbf{b}) clusters for different mean cluster sizes. Labels indicate the mean cluster size determined by polarization anisotropy measurements. Frequency offsets are $\omega_\mathrm{Ne} = 15035$\,cm$^{-1}$ for Ne and $\omega_\mathrm{Ar} = 14833$\,cm$^{-1}$ for Ar.}
    \label{fig:placeholder}
\end{figure*}

Fig\,\ref{fig:inhom-line_cluster_size} shows the inhomogeneous line profiles for different mean cluster sizes extracted from integration of 2D spectra along the anti-diagonal axis. 

\subsection{\label{sec:facet_size}Facet size estimation}
\begin{table}
\caption{\label{tab:cluster_size} Overview over the relationship between number of atoms and fully closed layers of icosahedral and cuboctahedral clusters.}
\begin{ruledtabular}
\begin{tabular}{lrrrrrrrr}
Fully closed layers&  3 & 4 & 5 & 6 & 7 & 8 & 9 & 10\\
Number of atoms & 147 & 309 & 561 & 923 & 1415 & 2057 & 2869 & 3871\\
\end{tabular}
\end{ruledtabular}
\end{table}
\begin{figure*}
    \centering
    \includegraphics[width = 0.8\linewidth]{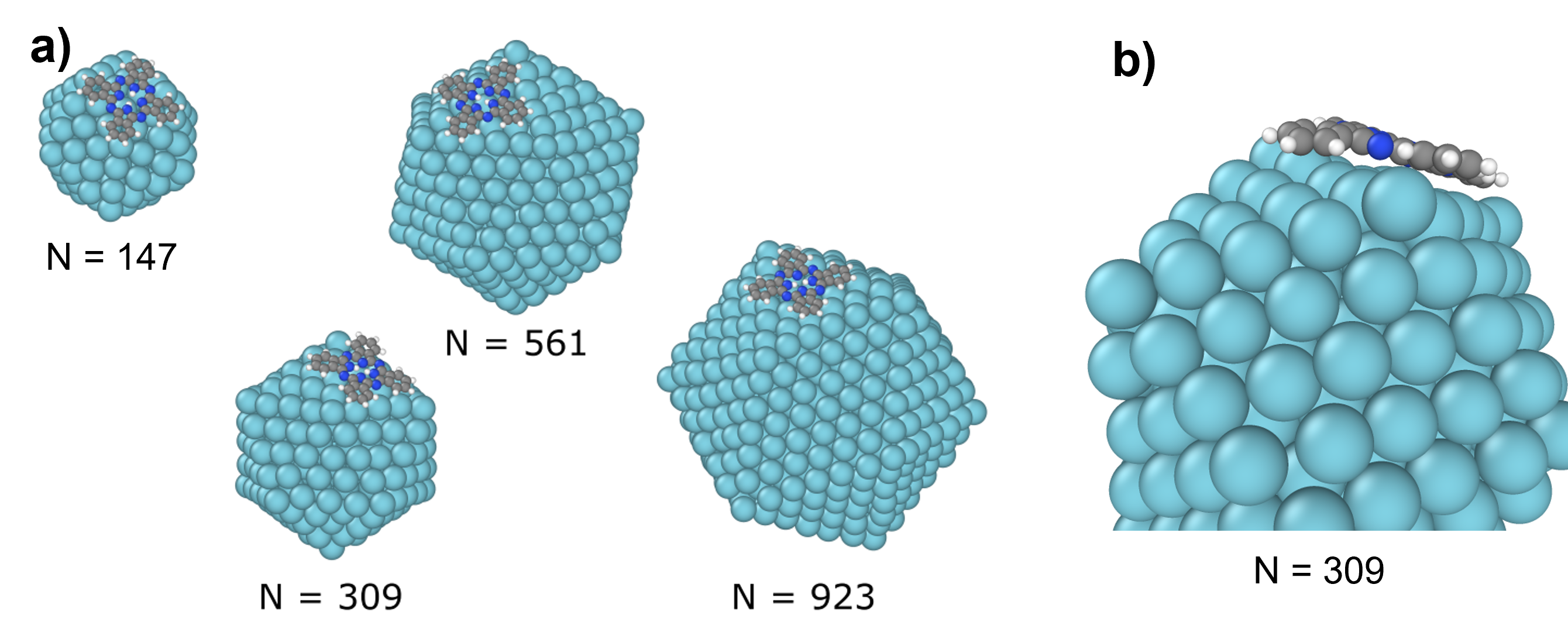}
    \caption{\label{fig:SI_Ar_images} Images of doped Ar clusters. (a) Closed-shell configurations used for the MD simulations of the Ar clusters. (b) Side view of cluster with size of 309 atoms. The cluster facet is too small to support the entire molecule which leads to a bending of the molecular structure at the edge of the facet.}
\end{figure*}
Tab.~\ref{tab:cluster_size} gives an overview over the relationship between the number of atoms in the cluster and the corresponding closed shells. 
The facet size $A$ of an icosahedral rare-gas cluster can be easily estimated based on some geometric considerations. 
The surface of an icosahedral cluster consists of equilateral triangles with spheres arranged in a hexagonal lattice and an edge length $a$, which is given by $N+1$ atoms for a cluster with $N$ shells \cite{Mackay1962}. Thus, $a$ can be estimated with $a = (N+1) \cdot 2r_\mathrm{vdW}$, where $r_\mathrm{vdW}$ is the van der Waals radius of the respective cluster atom species.
With this edge length the area of a facet can be calculated from the area of an equilateral triangle, i.e.
\begin{align}
    A_\mathrm{facet} = \dfrac{\sqrt{3}}{4} a^2.
\end{align}
With a van der Waals radius of $r_\mathrm{vdW}^\mathrm{Ar} = 1.88$\,$\mathrm{\AA}$ and $r_\mathrm{vdW}^\mathrm{Ne} = 1.54$\,$\mathrm{\AA}$, this yields a facet area of $A_\mathrm{Ar} = 98$\,$\mathrm{\AA}^2$ for an Ar cluster with $N = 3$ closed shells and $A_\mathrm{Ne} = 411$\,$\mathrm{\AA}^2$ for a Ne cluster with $N = 9$ closed shells. 

\section{Supplemental Material}\label{sec:sup_mat}
Supp. Material Video 1 shows a representative MD simulation of a Ne cluster with $N=2869$ atoms (9 closed shells) and an interaction potential of $V=1.5V_0$ and a trajectory from 0-1\,ns. 

Supp. Material Video 2 shows a representative MD simulation of a Ar cluster with $N=147$ atoms (3 closed shells) and an interaction potential of $V=V_0$ and a trajectory from 0-1\,ns. 

Supp. Material Video 3 shows a representative MD simulation of a Ar cluster with $N=147$ atoms (3 closed shells) and an interaction potential of $V=1.5V_0$ and a trajectory from 0-1\,ns.

\clearpage
\bibliography{Sources_Smart_Scanning, 2026DiffusionPaper}

@article{bangert_high-resolution_2022,
	title = {High-resolution two-dimensional electronic spectroscopy reveals the homogeneous line profile of chromophores solvated in nanoclusters},
	volume = {13},
	copyright = {2022 The Author(s)},
	issn = {2041-1723},
	url = {https://www.nature.com/articles/s41467-022-31021-z},
	doi = {10.1038/s41467-022-31021-z},
	language = {en},
	number = {1},
	urldate = {2022-06-13},
	journal = {Nat Commun},
	publisher = {Nature Publishing Group},
	author = {Bangert, Ulrich and Stienkemeier, Frank and Bruder, Lukas},
	month = jun,
	year = {2022},
	oldnote = {Number: 1},
	pages = {3350},
}

@article{sanda_center_2015,
	title = {Center {Line} {Slope} {Analysis} in {Two}-{Dimensional} {Electronic} {Spectroscopy}},
	volume = {119},
	issn = {1089-5639},
	url = {http://dx.doi.org/10.1021/acs.jpca.5b08909},
	doi = {10.1021/acs.jpca.5b08909},
	number = {44},
	urldate = {2017-04-07},
	journal = {J. Phys. Chem. A},
	author = {Šanda, František and Perlík, Václav and Lincoln, Craig N. and Hauer, Jürgen},
	month = nov,
	year = {2015},
	pages = {10893--10909},
}

@article{lewis_probing_2012,
	title = {Probing {Photosynthetic} {Energy} and {Charge} {Transfer} with {Two}-{Dimensional} {Electronic} {Spectroscopy}},
	volume = {3},
	issn = {1948-7185},
	url = {http://dx.doi.org/10.1021/jz201592v},
	doi = {10.1021/jz201592v},
	number = {4},
	urldate = {2013-12-02},
	journal = {J. Phys. Chem. Lett.},
	author = {Lewis, Kristin L. M. and Ogilvie, Jennifer P.},
	month = feb,
	year = {2012},
	pages = {503--510},
}

@article{brixner_two-dimensional_2005,
	title = {Two-dimensional spectroscopy of electronic couplings in photosynthesis},
	volume = {434},
	issn = {1476-4687},
	doi = {10.1038/nature03429},
	language = {eng},
	number = {7033},
	journal = {Nature},
	author = {Brixner, Tobias and Stenger, Jens and Vaswani, Harsha M and Cho, Minhaeng and Blankenship, Robert E and Fleming, Graham R},
	month = mar,
	year = {2005},
	pages = {625--628},
}

@article{lazonder_easy_2006,
	title = {Easy interpretation of optical two-dimensional correlation spectra},
	volume = {31},
	url = {http://ol.osa.org/abstract.cfm?URI=ol-31-22-3354},
	doi = {10.1364/OL.31.003354},
	number = {22},
	urldate = {2014-02-01},
	journal = {Opt. Lett.},
	author = {Lazonder, Kees and Pshenichnikov, Maxim S. and Wiersma, Douwe A.},
	month = nov,
	year = {2006},
	pages = {3354--3356},
}

@article{tekavec_fluorescence-detected_2007,
	title = {Fluorescence-detected two-dimensional electronic coherence spectroscopy by acousto-optic phase modulation},
	volume = {127},
	issn = {00219606},
	url = {http://jcp.aip.org/resource/1/jcpsa6/v127/i21/p214307_s1},
	doi = {doi:10.1063/1.2800560},
	number = {21},
	urldate = {2013-05-24},
	journal = {J. Chem. Phys.},
	author = {Tekavec, Patrick F. and Lott, Geoffrey A. and Marcus, Andrew H.},
	month = dec,
	year = {2007},
	pages = {214307--214307--21},
}

@article{barkai_theory_2004,
	title = {{THEORY} {OF} {SINGLE}-{MOLECULE} {SPECTROSCOPY}: {Beyond} the {Ensemble} {Average}},
	volume = {55},
	shorttitle = {{THEORY} {OF} {SINGLE}-{MOLECULE} {SPECTROSCOPY}},
	url = {http://dx.doi.org/10.1146/annurev.physchem.55.111803.143246},
	doi = {10.1146/annurev.physchem.55.111803.143246},
	number = {1},
	urldate = {2015-04-16},
	journal = {Annu. Rev. Phys. Chem.},
	author = {Barkai, Eli and Jung, YounJoon and Silbey, Robert},
	year = {2004},
	pages = {457--507},
}

@article{jonas_two-dimensional_2003,
	title = {Two-{Dimensional} {Femtosecond} {Spectroscopy}},
	volume = {54},
	url = {http://www.annualreviews.org/doi/abs/10.1146/annurev.physchem.54.011002.103907},
	doi = {10.1146/annurev.physchem.54.011002.103907},
	number = {1},
	urldate = {2013-05-23},
	journal = {Annu. Rev. Phys. Chem.},
	author = {Jonas, David M.},
	year = {2003},
	pages = {425--463},
}

@article{dvorak_spectroscopy_2012,
	title = {Spectroscopy of 3, 4, 9, 10-perylenetetracarboxylic dianhydride ({PTCDA}) attached to rare gas samples: {Clusters} vs. bulk matrices. {II}. {Fluorescence} emission spectroscopy},
	volume = {137},
	issn = {0021-9606, 1089-7690},
	shorttitle = {Spectroscopy of 3, 4, 9, 10-perylenetetracarboxylic dianhydride ({PTCDA}) attached to rare gas samples},
	url = {http://scitation.aip.org/content/aip/journal/jcp/137/16/10.1063/1.4759445},
	doi = {10.1063/1.4759445},
	number = {16},
	urldate = {2015-12-11},
	journal = {J Chem Phys},
	author = {Dvorak, Matthieu and Müller, Markus and Knoblauch, Tobias and Bünermann, Oliver and Rydlo, Alexandre and Minniberger, Stefan and Harbich, Wolfgang and Stienkemeier, Frank},
	month = oct,
	year = {2012},
	pages = {164302},
}

@article{fuller_experimental_2015,
	title = {Experimental {Implementations} of {Two}-{Dimensional} {Fourier} {Transform} {Electronic} {Spectroscopy}},
	volume = {66},
	url = {http://dx.doi.org/10.1146/annurev-physchem-040513-103623},
	doi = {10.1146/annurev-physchem-040513-103623},
	number = {1},
	urldate = {2015-04-16},
	journal = {Annu. Rev. Phys. Chem.},
	author = {Fuller, Franklin D. and Ogilvie, Jennifer P.},
	year = {2015},
	pages = {667--690},
}

@article{toennies_superfluid_2004,
	title = {Superfluid {Helium} {Droplets}: {A} {Uniquely} {Cold} {Nanomatrix} for {Molecules} and {Molecular} {Complexes}},
	volume = {43},
	issn = {1521-3773},
	shorttitle = {Superfluid {Helium} {Droplets}},
	url = {http://onlinelibrary.wiley.com/doi/10.1002/anie.200300611/abstract},
	doi = {10.1002/anie.200300611},
	language = {en},
	number = {20},
	urldate = {2013-05-24},
	journal = {Angew. Chem. Int. Ed.},
	author = {Toennies, J. Peter and Vilesov, Andrey F.},
	year = {2004},
	pages = {2622--2648},
}

@article{gutberlet_aggregation-induced_2009,
	title = {Aggregation-{Induced} {Dissociation} of {HCl}({H}$_{\textrm{2}}${O})$_{\textrm{4}}$ {Below} 1 {K}: {The} {Smallest} {Droplet} of {Acid}},
	volume = {324},
	copyright = {Copyright © 2009, American Association for the Advancement of Science},
	issn = {0036-8075, 1095-9203},
	shorttitle = {Aggregation-{Induced} {Dissociation} of {HCl}({H}$_{\textrm{2}}${O})$_{\textrm{4}}$ {Below} 1 {K}},
	url = {http://science.sciencemag.org/content/324/5934/1545},
	doi = {10.1126/science.1171753},
	language = {en},
	number = {5934},
	urldate = {2017-07-12},
	journal = {Science},
	author = {Gutberlet, Anna and Schwaab, Gerhard and Birer, {\"O}zg{\"u}r and Masia, Marco and Kaczmarek, Anna and Forbert, Harald and Havenith, Martina and Marx, Dominik},
	month = jun,
	year = {2009},
	pages = {1545--1548},
}

@article{zhang_surface-plasmon-driven_2018,
	title = {Surface-{Plasmon}-{Driven} {Hot} {Electron} {Photochemistry}},
	volume = {118},
	issn = {0009-2665},
	url = {https://doi.org/10.1021/acs.chemrev.7b00430},
	doi = {10.1021/acs.chemrev.7b00430},
	number = {6},
	urldate = {2019-01-24},
	journal = {Chem. Rev.},
	author = {Zhang, Yuchao and He, Shuai and Guo, Wenxiao and Hu, Yue and Huang, Jiawei and Mulcahy, Justin R. and Wei, Wei David},
	month = mar,
	year = {2018},
	pages = {2927--2954},
}

@article{moca_two-dimensional_2015,
	title = {Two-{Dimensional} {Electronic} {Spectroscopy} of {Chlorophyll} a: {Solvent} {Dependent} {Spectral} {Evolution}},
	volume = {119},
	issn = {1520-6106},
	shorttitle = {Two-{Dimensional} {Electronic} {Spectroscopy} of {Chlorophyll} a},
	url = {https://doi.org/10.1021/acs.jpcb.5b04339},
	doi = {10.1021/acs.jpcb.5b04339},
	number = {27},
	urldate = {2018-09-11},
	journal = {J. Phys. Chem. B},
	author = {Moca, Roberta and Meech, Stephen R. and Heisler, Ismael A.},
	month = jul,
	year = {2015},
	pages = {8623--8630},
}

@article{brinks_ultrafast_2014,
	title = {Ultrafast dynamics of single molecules},
	volume = {43},
	issn = {1460-4744},
	url = {https://pubs.rsc.org/en/content/articlelanding/2014/cs/c3cs60269a},
	doi = {10.1039/C3CS60269A},
	language = {en},
	number = {8},
	urldate = {2020-09-14},
	journal = {Chem. Soc. Rev.},
	publisher = {The Royal Society of Chemistry},
	author = {Brinks, Daan and Hildner, Richard and Dijk, Erik M. H. P. van and Stefani, Fernando D. and Nieder, Jana B. and Hernando, Jordi and Hulst, Niek F. van},
	month = mar,
	year = {2014},
	pages = {2476--2491},
}

@article{yoshida_anomalous_2020,
	title = {Anomalous {Temperature} {Dependence} of {Exciton} {Spectral} {Diffusion} in {Tetracene} {Thin} {Film}},
	volume = {11},
	url = {https://doi.org/10.1021/acs.jpclett.0c01537},
	doi = {10.1021/acs.jpclett.0c01537},
	number = {13},
	urldate = {2021-08-26},
	journal = {J. Phys. Chem. Lett.},
	publisher = {American Chemical Society},
	author = {Yoshida, Tatsuya and Watanabe, Kazuya and Petrović, Marin and Kralj, Marko},
	month = jul,
	year = {2020},
	pages = {5248--5254},
}

@article{mukamel_multidimensional_2000,
	title = {Multidimensional {Femtosecond} {Correlation} {Spectroscopies} of {Electronic} and {Vibrational} {Excitations}},
	volume = {51},
	url = {https://doi.org/10.1146/annurev.physchem.51.1.691},
	doi = {10.1146/annurev.physchem.51.1.691},
	number = {1},
	urldate = {2021-02-26},
	journal = {Annu. Rev. Phys. Chem.},
	author = {Mukamel, Shaul},
	year = {2000},
	oldnote = {\_eprint: https://doi.org/10.1146/annurev.physchem.51.1.691},
	pages = {691--729},
}

@article{roberts_characterization_2006,
	title = {Characterization of spectral diffusion from two-dimensional line shapes},
	volume = {125},
	issn = {0021-9606},
	url = {https://aip.scitation.org/doi/full/10.1063/1.2232271},
	doi = {10.1063/1.2232271},
	number = {8},
	urldate = {2021-12-13},
	journal = {J. Chem. Phys.},
	publisher = {American Institute of Physics},
	author = {Roberts, Sean T. and Loparo, Joseph J. and Tokmakoff, Andrei},
	month = aug,
	year = {2006},
	pages = {084502},
}

@article{dvorak_spectroscopy_2012-1,
	title = {Spectroscopy of 3, 4, 9, 10-perylenetetracarboxylic dianhydride ({PTCDA}) attached to rare gas samples: {Clusters} vs. bulk matrices. {I}. {Absorption} spectroscopy},
	volume = {137},
	issn = {0021-9606, 1089-7690},
	shorttitle = {Spectroscopy of 3, 4, 9, 10-perylenetetracarboxylic dianhydride ({PTCDA}) attached to rare gas samples},
	url = {http://scitation.aip.org/content/aip/journal/jcp/137/16/10.1063/1.4759443},
	doi = {10.1063/1.4759443},
	number = {16},
	urldate = {2015-12-11},
	journal = {J. Chem. Phys.},
	author = {Dvorak, Matthieu and Müller, Markus and Knoblauch, Tobias and Bünermann, Oliver and Rydlo, Alexandre and Minniberger, Stefan and Harbich, Wolfgang and Stienkemeier, Frank},
	month = oct,
	year = {2012},
	pages = {164301},
}

@article{murray_visible_2011,
	title = {Visible luminescence spectroscopy of free-base and zinc phthalocyanines isolated in cryogenic matrices},
	volume = {13},
	url = {https://pubs.rsc.org/en/content/articlelanding/2011/cp/c1cp22039j},
	doi = {10.1039/C1CP22039J},
	language = {en},
	number = {39},
	urldate = {2022-03-17},
	journal = {Phys. Chem. Chem. Phys.},
	publisher = {Royal Society of Chemistry},
	author = {Murray, Ciaran and Dozova, Nadia and G. McCaffrey, John and Shafizadeh, Niloufar and Chin, Wuthurath and Broquier, Michel and Crépin, Claudine},
	year = {2011},
	pages = {17543--17554},
}

@article{geissinger_hole-burning_1992,
	title = {Hole-burning in dye-doped rare-gas matrices},
	volume = {197},
	issn = {0009-2614},
	url = {https://www.sciencedirect.com/science/article/pii/000926149286043H},
	doi = {10.1016/0009-2614(92)86043-H},
	language = {en},
	number = {1},
	urldate = {2022-03-17},
	journal = {Chemical Physics Letters},
	author = {Geissinger, P. and Haarer, D.},
	month = sep,
	year = {1992},
	pages = {175--180},
}

@article{briant_reaction_2022,
	title = {Reaction dynamics within a cluster environment},
	volume = {24},
	url = {https://pubs.rsc.org/en/content/articlelanding/2022/cp/d1cp05783a},
	doi = {10.1039/D1CP05783A},
	language = {en},
	number = {17},
	urldate = {2023-11-10},
	journal = {Phys. Chem. Chem. Phys.},
	publisher = {Royal Society of Chemistry},
	author = {Briant, Marc and Mestdagh, Jean-Michel and Gaveau, Marc-André and Poisson, Lionel},
	year = {2022},
	pages = {9807--9835},
}

@article{schwarz_menage-a-trois_2017,
	title = {Ménage-à-trois: single-atom catalysis, mass spectrometry, and computational chemistry},
	volume = {7},
	issn = {2044-4753, 2044-4761},
	shorttitle = {Ménage-à-trois},
	url = {https://xlink.rsc.org/?DOI=C6CY02658C},
	doi = {10.1039/C6CY02658C},
	language = {en},
	number = {19},
	urldate = {2025-12-11},
	journal = {Catal. Sci. Technol.},
	author = {Schwarz, Helmut},
	year = {2017},
	pages = {4302--4314},
}

@article{hansen_we_2004,
	series = {Special {Issue}: {In} honour of {Tilmann} {Mark}},
	title = {Do we know the value of the {Gspann} parameter?},
	volume = {233},
	issn = {1387-3806},
	url = {https://www.sciencedirect.com/science/article/pii/S1387380604000557},
	doi = {10.1016/j.ijms.2003.12.021},
	number = {1},
	urldate = {2026-01-27},
	journal = {International Journal of Mass Spectrometry},
	author = {Hansen, K. and Campbell, E. E. B.},
	month = apr,
	year = {2004},
	pages = {215--221},
}

@article{awali_time-resolved_2021,
	title = {Time-{Resolved} {Observation} of the {Solvation} {Dynamics} of a {Rydberg} {Excited} {Molecule} {Deposited} on an {Argon} {Cluster}. {II}. {DABCO}${^\ensuremath{\star}}$ at {Long} {Time} {Delays}},
	volume = {125},
	issn = {1089-5639},
	url = {https://doi.org/10.1021/acs.jpca.1c01942},
	doi = {10.1021/acs.jpca.1c01942},
	number = {20},
	urldate = {2026-01-27},
	journal = {J. Phys. Chem. A},
	publisher = {American Chemical Society},
	author = {Awali, Slim and Mestdagh, Jean-Michel and Gaveau, Marc-André and Briant, Marc and Soep, Benoît and Mazet, Vincent and Poisson, Lionel},
	month = may,
	year = {2021},
	pages = {4341--4351},
}

@article{do_measuring_2019,
	title = {Measuring {Ultrafast} {Spectral} {Diffusion} and {Correlation} {Dynamics} by {Two}-{Dimensional} {Electronic} {Spectroscopy}},
	volume = {14},
	copyright = {© 2019 Wiley-VCH Verlag GmbH \& Co. KGaA, Weinheim},
	issn = {1861-471X},
	url = {https://onlinelibrary.wiley.com/doi/abs/10.1002/asia.201900994},
	doi = {10.1002/asia.201900994},
	language = {en},
	number = {22},
	urldate = {2026-01-27},
	journal = {Chem. Asian J.},
	author = {Do, Thanh Nhut and Khyasudeen, M. Faisal and Nowakowski, Paweł J. and Zhang, Zhengyang and Tan, Howe-Siang},
	year = {2019},
	oldnote = {\_eprint: https://aces.onlinelibrary.wiley.com/doi/pdf/10.1002/asia.201900994},
	pages = {3992--4000},
}

@article{maly_wavelike_2020,
	title = {From wavelike to sub-diffusive motion: exciton dynamics and interaction in squaraine copolymers of varying length},
	volume = {11},
	issn = {2041-6539},
	shorttitle = {From wavelike to sub-diffusive motion},
	url = {https://pubs.rsc.org/en/content/articlelanding/2020/sc/c9sc04367e},
	doi = {10.1039/C9SC04367E},
	language = {en},
	number = {2},
	urldate = {2026-01-27},
	journal = {Chem. Sci.},
	publisher = {The Royal Society of Chemistry},
	author = {Malý, Pavel and Lüttig, Julian and Turkin, Arthur and Dostál, Jakub and Lambert, Christoph and Brixner, Tobias},
	month = jan,
	year = {2020},
	pages = {456--466},
}

@article{cocker_tracking_2016,
	title = {Tracking the ultrafast motion of a single molecule by femtosecond orbital imaging},
	volume = {539},
	copyright = {2016 Macmillan Publishers Limited, part of Springer Nature. All rights reserved.},
	issn = {1476-4687},
	url = {https://www.nature.com/articles/nature19816},
	doi = {10.1038/nature19816},
	language = {en},
	number = {7628},
	urldate = {2026-01-27},
	journal = {Nature},
	publisher = {Nature Publishing Group},
	author = {Cocker, Tyler L. and Peller, Dominik and Yu, Ping and Repp, Jascha and Huber, Rupert},
	month = nov,
	year = {2016},
	pages = {263--267},
}

@article{farnik_pickup_2021,
	title = {Pickup and reactions of molecules on clusters relevant for atmospheric and interstellar processes},
	volume = {23},
	issn = {1463-9084},
	url = {https://pubs.rsc.org/en/content/articlelanding/2021/cp/d0cp06127a},
	doi = {10.1039/D0CP06127A},
	language = {en},
	number = {5},
	urldate = {2026-03-20},
	journal = {Phys. Chem. Chem. Phys.},
	publisher = {The Royal Society of Chemistry},
	author = {Fárník, Michal and Fedor, Juraj and Kočišek, Jaroslav and Lengyel, Jozef and Pluhařová, Eva and Poterya, Viktoriya and Pysanenko, Andriy},
	month = feb,
	year = {2021},
	pages = {3195--3213},
}

@article{potapov_physics_2021,
	title = {Physics and chemistry on the surface of cosmic dust grains: a laboratory view},
	volume = {40},
	issn = {0144-235X},
	shorttitle = {Physics and chemistry on the surface of cosmic dust grains},
	url = {https://doi.org/10.1080/0144235X.2021.1918498},
	doi = {10.1080/0144235X.2021.1918498},
	number = {2},
	urldate = {2026-03-20},
	journal = {Int. Rev. Phys. Chem.},
	publisher = {Taylor \& Francis},
	author = {Potapov, Alexey and McCoustra, Martin},
	month = apr,
	year = {2021},
	oldnote = {\_eprint: https://doi.org/10.1080/0144235X.2021.1918498},
	pages = {299--364},
}

@article{mestdagh_cluster_1997,
	title = {Cluster isolated chemical reactions},
	volume = {16},
	issn = {0144-235X},
	url = {https://doi.org/10.1080/014423597230280},
	doi = {10.1080/014423597230280},
	number = {2},
	urldate = {2026-03-20},
	journal = {Int. Rev. Phys. Chem.},
	publisher = {Taylor \& Francis},
	author = {Mestdagh, J.M. and Gaveau, M.A. and Gee, C. and Sublemontier, O. and Visticot, J.P.},
	month = apr,
	year = {1997},
	oldnote = {\_eprint: https://doi.org/10.1080/014423597230280},
	pages = {215--247},
}

@article{briant_cluster_2000,
	title = {Cluster isolated chemical reaction ({CICR}) spectroscopy: {Ba} atoms and {Ba}({CH4})n complexes on large neon clusters},
	volume = {112},
	issn = {0021-9606},
	shorttitle = {Cluster isolated chemical reaction ({CICR}) spectroscopy},
	url = {https://doi.org/10.1063/1.480820},
	doi = {10.1063/1.480820},
	number = {4},
	urldate = {2026-03-20},
	journal = {J. Chem. Phys.},
	author = {Briant, M. and Gaveau, M. A. and Mestdagh, J. M. and Visticot, J. P.},
	month = jan,
	year = {2000},
	pages = {1744--1756},
}

@article{thon_wco6_2017,
	series = {Selected {Papers} from the 19th {International} {Conference} on {Dynamical} {Processes} in {Excited} {States} of {Solids} ({DPC}’16)},
	title = {W({CO})6 in cryogenic solids: {A} comparative study of vibrational properties},
	volume = {191},
	issn = {0022-2313},
	shorttitle = {W({CO})6 in cryogenic solids},
	url = {https://www.sciencedirect.com/science/article/pii/S0022231316314491},
	doi = {10.1016/j.jlumin.2017.01.041},
	urldate = {2026-03-25},
	journal = {Journal of Luminescence},
	author = {Thon, Raphaël and Chin, Wutharath and Chamma, Didier and Gutiérrez-Quintanilla, Alejandro and Chevalier, Michèle and Galaup, Jean-Pierre and Crépin, Claudine},
	month = nov,
	year = {2017},
	pages = {78--86},
}

@article{arabei_stimulated_2015,
	title = {Stimulated emission in cryogenic samples doped with free-base tetraazaporphine},
	volume = {17},
	url = {https://pubs.rsc.org/en/content/articlelanding/2015/cp/c5cp01286d},
	doi = {10.1039/C5CP01286D},
	language = {en},
	number = {22},
	urldate = {2026-03-25},
	journal = {Phys. Chem. Chem. Phys.},
	publisher = {Royal Society of Chemistry},
	author = {Arabei, Serguei and G. McCaffrey, John and Galaup, Jean-Pierre and Shafizadeh, Niloufar and Crépin, Claudine},
	year = {2015},
	pages = {14931--14942},
}

@article{jiang_ultrafast_2024,
	title = {Ultrafast photoinduced {C}-{H} bond formation from two small inorganic molecules},
	volume = {15},
	copyright = {2024 The Author(s)},
	issn = {2041-1723},
	url = {https://www.nature.com/articles/s41467-024-47137-3},
	doi = {10.1038/s41467-024-47137-3},
	language = {en},
	number = {1},
	urldate = {2026-06-23},
	journal = {Nat Commun},
	publisher = {Nature Publishing Group},
	author = {Jiang, Zhejun and Huang, Hao and Lu, Chenxu and Zhou, Lianrong and Pan, Shengzhe and Qiang, Junjie and Shi, Menghang and Ye, Zhengjun and Lu, Peifen and Ni, Hongcheng and Zhang, Wenbin and Wu, Jian},
	month = apr,
	year = {2024},
	pages = {2854},
}

@article{zhou_ultrafast_2023,
	title = {Ultrafast formation dynamics of {D3}+ from the light-driven bimolecular reaction of the {D2}–{D2} dimer},
	volume = {15},
	copyright = {2023 The Author(s), under exclusive licence to Springer Nature Limited},
	issn = {1755-4349},
	url = {https://www.nature.com/articles/s41557-023-01230-0},
	doi = {10.1038/s41557-023-01230-0},
	language = {en},
	number = {9},
	urldate = {2026-06-23},
	journal = {Nat. Chem.},
	publisher = {Nature Publishing Group},
	author = {Zhou, Lianrong and Ni, Hongcheng and Jiang, Zhejun and Qiang, Junjie and Jiang, Wenyu and Zhang, Wenbin and Lu, Peifen and Wen, Jin and Lin, Kang and Zhu, Meifang and Dörner, Reinhard and Wu, Jian},
	month = sep,
	year = {2023},
	pages = {1229--1235},
}

@incollection{de_boer_chapter_1957,
	title = {Chapter {I} {Quantum} {Effects} and {Exchange} {Effects} on the {Thermodynamic} {Properties} of {Liquid} {Helium}},
	volume = {2},
	copyright = {https://www.elsevier.com/tdm/userlicense/1.0/},
	isbn = {978-0-444-53308-1},
	url = {https://linkinghub.elsevier.com/retrieve/pii/S0079641708601006},
	doi = {10.1016/S0079-6417(08)60100-6},
	oldlanguage = {en},
	urldate = {2026-06-23},
	booktitle = {Progress in {Low} {Temperature} {Physics}},
	publisher = {Elsevier},
	author = {de Boer, J},
	year = {1957},
	pages = {1--58},
}

@article{maiuri_ultrafast_2020,
	title = {Ultrafast {Spectroscopy}: {State} of the {Art} and {Open} {Challenges}},
	volume = {142},
	issn = {0002-7863},
	shorttitle = {Ultrafast {Spectroscopy}},
	url = {https://doi.org/10.1021/jacs.9b10533},
	doi = {10.1021/jacs.9b10533},
	number = {1},
	urldate = {2026-06-23},
	journal = {J. Am. Chem. Soc.},
	publisher = {American Chemical Society},
	author = {Maiuri, Margherita and Garavelli, Marco and Cerullo, Giulio},
	month = jan,
	year = {2020},
	pages = {3--15},
}

@article{klimek_hierarchical_2026,
	title = {Hierarchical friction memory leads to subdiffusive configurational dynamics of fast-folding proteins},
	volume = {123},
	url = {https://www.pnas.org/doi/abs/10.1073/pnas.2516506123},
	doi = {10.1073/pnas.2516506123},
	number = {6},
	urldate = {2026-06-23},
	journal = {PNAS},
	publisher = {Proceedings of the National Academy of Sciences},
	author = {Klimek, Anton and Dalton, Benjamin A. and Tepper, Lucas and Netz, Roland R.},
	month = feb,
	year = {2026},
	pages = {e2516506123},
}

@article{fried_structure_1992,
	title = {Structure, dynamics, and the electronic absorption of benzene–argon clusters},
	volume = {96},
	issn = {0021-9606},
	url = {https://doi.org/10.1063/1.462501},
	doi = {10.1063/1.462501},
	number = {1},
	urldate = {2026-06-25},
	journal = {J. Chem. Phys.},
	author = {Fried, Laurence E. and Mukamel, Shaul},
	month = jan,
	year = {1992},
	pages = {116--135},
}

@article{gaigeot_diffusion_1997,
	title = {Diffusion and clustering of {N2O} molecules in argon clusters: {A} theoretical approach by molecular dynamics simulations},
	volume = {106},
	issn = {0021-9606},
	shorttitle = {Diffusion and clustering of {N2O} molecules in argon clusters},
	url = {https://doi.org/10.1063/1.474020},
	doi = {10.1063/1.474020},
	number = {22},
	urldate = {2026-06-25},
	journal = {J. Chem. Phys.},
	author = {Gaigeot, M.-P. and de Pujo, P. and Brenner, V. and Millié, Ph.},
	month = jun,
	year = {1997},
	pages = {9155--9171},
}

@article{sano_theory_1981,
	title = {Theory of diffusion-controlled reactions on spherical surfaces and its application to reactions on micellar surfaces},
	volume = {75},
	issn = {0021-9606, 1089-7690},
	url = {https://pubs.aip.org/jcp/article/75/6/2870/791868/Theory-of-diffusion-controlled-reactions-on},
	doi = {10.1063/1.442360},
	language = {en},
	number = {6},
	urldate = {2026-06-25},
	journal = {J Chem Phys},
	author = {Sano, Hisatake and Tachiya, M.},
	month = sep,
	year = {1981},
	pages = {2870--2878},
}

@article{berry_clusters_2013,
	series = {Clusters: {From} {Dimers} to {Nanoparticles}},
	title = {Clusters as tools to link macro and micro approaches},
	volume = {1021},
	issn = {2210-271X},
	url = {https://www.sciencedirect.com/science/article/pii/S2210271X13002387},
	doi = {10.1016/j.comptc.2013.06.008},
	urldate = {2026-06-25},
	journal = {Computational and Theoretical Chemistry},
	author = {Berry, R. Stephen and Smirnov, Boris M.},
	month = oct,
	year = {2013},
	pages = {2--6},
}

@incollection{colombo_imaging_2023,
	title = {Imaging {Clusters} and {Their} {Dynamics} with {Single}-shot {Coherent} {Diffraction}},
	volume = {25},
	isbn = {978-1-83767-114-4},
	url = {https://doi.org/10.1039/BK9781837671564-00172},
	doi = {10.1039/BK9781837671564-00172},
	urldate = {2026-06-25},
	booktitle = {Structural {Dynamics} with {X}-ray and {Electron} {Scattering}},
	publisher = {Royal Society of Chemistry},
	author = {Colombo, Alessandro and Rupp, Daniela},
	editor = {Amini, Kasra and Rouzée, Arnaud and Vrakking, Marc J J},
	month = dec,
	year = {2023},
	pages = {0},
}

@misc{michaelcomm,
  author       = {Walter, Michael},
  title        = {Private communication},
  year         = {2025},
}

@PREAMBLE{
 "\providecommand{\noopsort}[1]{}" 
 # "\providecommand{\singleletter}[1]{#1}%" 
}

@ARTICLE{Farges1980,
    author  = {Farges, J. and de Feraudy, M. F. and Raoult, B. and Torchet, G.}, 
    title   = {Structure and temperature of rare gas clusters in a supersonic expansion},
    year    = {1980}, 
    journal = {Surf. Sci.}, 
    volume  = {106},
    number  = {1},
    pages   = {95-100},
    issn    = {0039-6028},
    doi     = {https://doi.org/10.1016/0039-6028(81)90186-2},
    url     = {https://www.sciencedirect.com/science/article/pii/0039602881901862}
}

@article{Bell2015,
    author      = {Bell, J. D. and Conrad, R. and Siemens, M. E.},
    journal     = {Opt. Lett.},
    number      = {7},
    pages       = {1157--1160},
    publisher   = {Optica Publishing Group},
    title       = {Analytical calculation of two-dimensional spectra},
    volume      = {40},
    month       = {Apr},
    year        = {2015},
    url         = {https://opg.optica.org/ol/abstract.cfm?URI=ol-40-7-1157},
    doi         = {10.1364/OL.40.001157}
}

@inbook{Yuste2008,
    author      = {Yuste, Santos Bravo and Lindenberg, Katja and Ruiz-Lorenzo, Juan Jesús},
    publisher   = {John Wiley \& Sons, Ltd},
    isbn        = {9783527622979},
    title       = {Subdiffusion-Limited Reactions},
    booktitle   = {Anomalous Transport},
    chapter     = {13},
    pages       = {367-395},
    doi         = {https://doi.org/10.1002/9783527622979.ch13},
    url         = {https://onlinelibrary.wiley.com/doi/abs/10.1002/9783527622979.ch13},
    oldeprint      = {https://onlinelibrary.wiley.com/doi/pdf/10.1002/9783527622979.ch13},
    year        = {2008},
}

@book{Avraham2000,
    place       = {Cambridge},
    title       = {Beyond random walks},
    booktitle   = {Diffusion and Reactions in Fractals and Disordered Systems}, 
    publisher   = {Cambridge University Press}, 
    author      = {ben-Avraham, Daniel and Havlin, Shlomo},
    year        = {2000},
    pages       = {46-48},
    doi         = {https://doi.org/10.1017/CBO9780511605826}}

@article{Awali2014,
author ="Awali, Slim and Poisson, Lionel and Soep, Benoît and Gaveau, Marc-André and Briant, Marc and Pothier, Christophe and Mestdagh, Jean-Michel and Rhouma, Mounir Ben El Hadj and Hochlaf, Majdi and Mazet, Vincent and Faisan, Sylvain",
title  ="Time resolved observation of the solvation dynamics of a Rydberg excited molecule deposited on an argon cluster-I: DABCO at short times",
journal  ="Phys. Chem. Chem. Phys.",
year  ="2014",
volume  ="16",
issue  ="2",
pages  ="516-526",
publisher  ="The Royal Society of Chemistry",
doi  ="10.1039/C3CP53172D",
url  ="http://dx.doi.org/10.1039/C3CP53172D",
}

@article{Bruder2019,
doi = {10.1088/1361-6455/ab319f},
url = {https://dx.doi.org/10.1088/1361-6455/ab319f},
year = {2019},
month = {aug},
publisher = {IOP Publishing},
volume = {52},
number = {18},
pages = {183501},
author = {Lukas Bruder and Ulrich Bangert and Marcel Binz and Daniel Uhl and Frank Stienkemeier},
title = {Coherent multidimensional spectroscopy in the gas phase},
journal = {Journal of Physics B: Atomic, Molecular and Optical Physics},
}

@article{heinz2013thermodynamically,
  title={Thermodynamically consistent force fields for the assembly of inorganic, organic, and biological nanostructures: the INTERFACE force field},
  author={Heinz, Hendrik and Lin, Tzu-Jen and Kishore Mishra, Ratan and Emami, Fateme S},
  journal={Langmuir},
  volume={29},
  number={6},
  pages={1754--1765},
  year={2013},
  publisher={ACS Publications}
}

@article{pramanik2017carbon,
  title={Carbon nanotube dispersion in solvents and polymer solutions: mechanisms, assembly, and preferences},
  author={Pramanik, Chandrani and Gissinger, Jacob R and Kumar, Satish and Heinz, Hendrik},
  journal={ACS nano},
  volume={11},
  number={12},
  pages={12805--12816},
  year={2017},
  publisher={ACS Publications}
}

@misc{heinzWeb2021,
  title        = {Interfaces Laboratory},
  howpublished = {\protect\url{https://bionanostructures.com/interface-md/}},
  note         = {Accessed: 2021-11-16}
}

@article{Hagena1972,
    author = {Hagena, O. F. and Obert, W.},
    title = {Cluster Formation in Expanding Supersonic Jets: Effect of Pressure, Temperature, Nozzle Size, and Test Gas},
    journal = {J. Chem. Phys.},
    volume = {56},
    number = {5},
    pages = {1793-1802},
    year = {1972},
    month = {03},
    issn = {0021-9606},
    doi = {10.1063/1.1677455},
    url = {https://doi.org/10.1063/1.1677455},
    oldeprint = {https://pubs.aip.org/aip/jcp/article-pdf/56/5/1793/18879085/1793\_1\_online.pdf},
}

@book{HammZanni2011,
	title = {Concepts and \textsc{M}ethods of 2\textsc{D} \textsc{I}nfrared \textsc{S}pectroscopy},
	author = {Hamm, Peter and Zanni, Martin},
	isbn = {978-1-107-00005-6},
	year = {2011},
	publisher = {Cambridge University Press},
	edition = {1},
}

@article{Lomsadze2018,
	author={Lomsadze, Bachana
	and Smith, Brad C.
	and Cundiff, Steven T.},
	title={Tri-comb spectroscopy},
	journal={Nat. Photonics},
	year={2018},
	month={Nov},
	day={01},
	volume={12},
	number={11},
	pages={676-680},
	issn={1749-4893},
	doi={10.1038/s41566-018-0267-4},
	url={https://doi.org/10.1038/s41566-018-0267-4}
}

@article{Hagena1981,
	title = {Nucleation and growth of clusters in expanding nozzle flows},
	journal = {Surf. Sci.},
	volume = {106},
	number = {1},
	pages = {101-116},
	year = {1981},
	issn = {0039-6028},
	doi = {https://doi.org/10.1016/0039-6028(81)90187-4},
	url = {https://www.sciencedirect.com/science/article/pii/0039602881901874},
	author = {Otto F. Hagena},
}

@article{nose1984unified,
  title={A unified formulation of the constant temperature molecular dynamics methods},
  author={Nos{\'e}, Shuichi},
  journal={The Journal of chemical physics},
  volume={81},
  number={1},
  pages={511--519},
  year={1984},
  publisher={AIP Publishing}
}

@article{hoover1985canonical,
  title={Canonical dynamics: Equilibrium phase-space distributions},
  author={Hoover, William G},
  journal={Physical review A},
  volume={31},
  number={3},
  pages={1695},
  year={1985},
  publisher={APS}
}

@article{elsasser2024structural,
  title={Structural transition in the single layer growth of diindenoperylene on silica},
  author={Els{\"a}sser, Philipp and Schilling, Tanja},
  journal={The Journal of Chemical Physics},
  volume={161},
  number={9},
  year={2024},
  publisher={AIP Publishing}
}

@article{elsasser2023optimizing,
  title={Optimizing the structure of acene clusters},
  author={Els{\"a}sser, Philipp and Schilling, Tanja},
  journal={The Journal of Chemical Physics},
  volume={158},
  number={12},
  year={2023},
  publisher={AIP Publishing}
}

@article{patil2025high,
  title={High-performance, multi-component epoxy resin simulation for predicting thermo-mechanical property evolution during curing},
  author={Patil, Sagar Umesh and Kemppainen, Josh and Maiaru, Marianna and Odegard, Gregory M},
  journal={Polymer Journal},
  volume={57},
  number={5},
  pages={539--552},
  year={2025},
  publisher={Nature Publishing Group UK London}
}

@article{deshpande2021prediction,
  title={Prediction of the interfacial properties of high-performance polymers and flattened CNT-reinforced composites using molecular dynamics},
  author={Deshpande, Prathamesh P and Radue, Matthew S and Gaikwad, Prashik and Bamane, Swapnil and Patil, Sagar U and Pisani, William A and Odegard, Gregory M},
  journal={Langmuir},
  volume={37},
  number={39},
  pages={11526--11534},
  year={2021},
  publisher={ACS Publications}
}

@article{vsiber2004vibrations,
  title={Vibrations of closed-shell Lennard-Jones icosahedral and cuboctahedral clusters and their effect on the cluster ground-state energy},
  author={{\v{S}}iber, Antonio},
  journal={Physical Review B—Condensed Matter and Materials Physics},
  volume={70},
  number={7},
  pages={075407},
  year={2004},
  publisher={APS}
}

@article{elsasser1987energetics,
  title={Energetics of Al13, Cu13 and Ni13 clusters of various symmetries},
  author={Elsasser, C and Fahnle, M and Brandt, EH and Bohm, MC},
  journal={Journal of Physics F: Metal Physics},
  volume={17},
  number={11},
  pages={L301--L304},
  year={1987}
}

@article{Mackay1962,
author = "Mackay, A. L.",
title = "{A dense non-crystallographic packing of equal spheres}",
journal = "Acta Crystallographica",
year = "1962",
volume = "15",
number = "9",
pages = "916--918",
month = "Sep",
doi = {10.1107/S0365110X6200239X},
url = {https://doi.org/10.1107/S0365110X6200239X},
}

\end{document}